\documentclass[prd,showpacs,superscriptaddress,nofootinbib,floatfix,11pt]{revtex4-2}

\usepackage{amsmath}
\usepackage[scr=boondox]{mathalpha}
\usepackage{array}
\usepackage{mathtools}
\usepackage{amssymb}
\usepackage{wasysym}
\usepackage{color}
\usepackage{slashed}
\usepackage{bbm}
\usepackage{graphicx}
\usepackage[caption=false]{subfig}
\usepackage{svg}
\usepackage{xcolor}
\definecolor{Tiffany}{rgb}{0, 0.7, 0.7}
\usepackage[colorlinks=true,
breaklinks=true,
urlcolor=magenta,
linkcolor=magenta,
citecolor=Tiffany]{hyperref}
\usepackage{tikz}
\usetikzlibrary{positioning, arrows.meta, calc}
\usepackage{subfig}
\usepackage{multirow}
\usepackage{booktabs}
\usepackage{orcidlink}
\usepackage[normalem]{ulem}
\renewcommand{\bm}[1]{\boldsymbol{#1}}
\usepackage{braket}

\allowdisplaybreaks

\newcommand{\ucas}{\affiliation{School of Physical Sciences, University of Chinese Academy of Sciences, Beijing 100049, China}}
\newcommand{\itp}{\affiliation{Institute of Theoretical Physics, Chinese Academy of Sciences, Beijing 100190, China}}
\newcommand{\tmu}{\affiliation{Department of Physics, Tokyo Metropolitan University, Hachioji 192-0397, Japan}}
\newcommand{\argonne}{\affiliation{High Energy Physics Division, Argonne National Laboratory, Argonne, IL 60439, USA}}
\newcommand{\nwu}{\affiliation{Department of Physics and Astronomy, Northwestern University, Evanston, IL 60208, USA}}

\begin{document}

\title{Entanglement Suppression, Quantum Statistics and Symmetries\\in Spin-3/2 Baryon Scatterings}

\author{\mbox{Tao-Ran Hu}\orcidlink{0009-0003-9720-0171}}\email{hutaoran21@mails.ucas.ac.cn}
\ucas

\author{\mbox{Katsuyoshi Sone}\orcidlink{0000-0002-2755-3284}}\email{sone-katsuyoshi@ed.tmu.ac.jp}
\tmu

\author{\mbox{Feng-Kun Guo}\orcidlink{0000-0002-2919-2064}}\email{fkguo@itp.ac.cn}
\itp\ucas

\author{\mbox{Tetsuo Hyodo}\orcidlink{0000-0002-4145-9817}}\email{hyodo@rcnp.osaka-u.ac.jp}\thanks{\\Present address: Research Center for Nuclear Physics (RCNP), Ibaraki, Osaka 567-0047, Japan}
\tmu

\author{\mbox{Ian Low}\orcidlink{0000-0002-7570-9597}}\email{ilow@northwestern.edu}
\argonne\nwu

\date{\today}

\begin{abstract}

We explore the interplay among entanglement suppression, quantum statistics and enhanced symmetries in the non-relativistic $S$-wave scattering involving the lowest-lying spin-3/2 baryons, which can be considered as four-dimensional qudits. These baryons form a ten-dimensional representation (decuplet) under the SU(3) light-flavor symmetry and, in this limit, are considered indistinguishable under strong interactions. Treating the $S$-matrix in the spin-3/2 baryon-baryon scattering as a quantum logic gate in the spin space, we study the consequence of entanglement suppression and compute the entanglement power of the $S$-matrix. When the entanglement power vanishes, the $S$-matrix is either an Identity or a SWAP gate and spin-flavor symmetries and/or non-relativistic conformal invariance emerge, as previously observed in spin-1/2 baryons. In the case of scattering identical particles,  the entanglement power never vanishes due to constraints from spin statistics, which we interpret as projection-valued measurements onto symmetric or antisymmetric Hilbert space and define the entanglement power accordingly. When the entanglement power is non-vanishing but sits at a global or local minimum, enhanced symmetries still emerge and the $S$-matrix can be interpreted as an Identity or a SWAP gate acting on the restricted Hilbert space allowed by quantum statistics. In general, when scattering identical spin-$s$ particles, we identify an enhanced SU(2$s$+1)$_{\rm spin}$ symmetry for the Identity gate.

\end{abstract}

\maketitle

\newpage

\tableofcontents

\newpage

\section{Introduction}

One remaining fundamental challenge in the Standard Model of particle physics is the non-perturbative dynamics of the strong interaction. Since all quarks and gluons are confined inside color-singlet hadrons, in order to gain more insights, one needs to study hadron properties and hadron-hadron interactions to reveal patterns from either experimental data or lattice quantum chromodynamics (QCD) calculations. For instance, let us consider the nucleon-nucleon interaction at low energies. A pair of nucleons, being spin-1/2 and isospin-1/2 particles, in $S$ waves can form either an isospin-singlet $^3S_1$ or an isospin-triplet $^1S_0$ combination. In principle, the interactions in these pairs can be very much different. However, from experimental data on nuclear force, the interaction strengths in these pairs are approximately equal.\footnote{The deuteron is a $^3S_1$ bound state with a binding energy of 2.2~MeV, while the $^1S_0$ $nn$ pair has a virtual state about 66~keV below threshold (see, e.g., Ref.~\cite{Wiringa:1994wb}). Were the attraction in the latter pair a little bit stronger, a bound state would also be formed. As a consequence of the existence of the shallow bound or virtual state, both channels have unnaturally large scattering lengths.} This was traditionally explained by considering the large $N_c$, with $N_c$ the number of colors in QCD, limit~\cite{Kaplan:1995yg}; however, recently, a new explanation using the so-called entanglement suppression was proposed~\cite{Beane:2018oxh}. It is thus interesting to explore whether the entanglement-suppression hypothesis, that minimizing the entanglement in a quantum scattering process leads to an enlarged symmetry absent in the original action of the theory, is universal in the infrared region of QCD.

Quantum entanglement is the most prominent feature of quantum mechanics. Yet its significance in the context of particle scatterings and quantum field theory was not greatly appreciated until recently~\cite{Cervera-Lierta:2017tdt, Afik:2020onf, Aoude:2020mlg, Aoude:2023hxv, Sakurai:2023nsc, Barr:2024djo, Aoude:2024xpx,Low:2024mrk, Low:2024hvn, Blasone:2025tor}. One of the most fascinating results is the observation that entanglement suppression, or maximization, in 2-to-2 scattering processes could lead to enhanced symmetries in scattering dynamics~\cite{Beane:2018oxh, Low:2021ufv, Liu:2022grf, Carena:2023vjc, Liu:2023bnr, Hu:2024hex, Kowalska:2024kbs, Chang:2024wrx, McGinnis:2025brt, Carena:2025wyh}. Other intriguing developments include using entanglement to understand the flavor patterns in the Standard Model~\cite{Thaler:2024anb} or the dynamics of electroweak phase transition~\cite{Liu:2025pny}. These studies open up a new frontier to understand fundamental interactions of Nature from the information-theoretic perspective and offer new insights into observables in high-energy colliders.

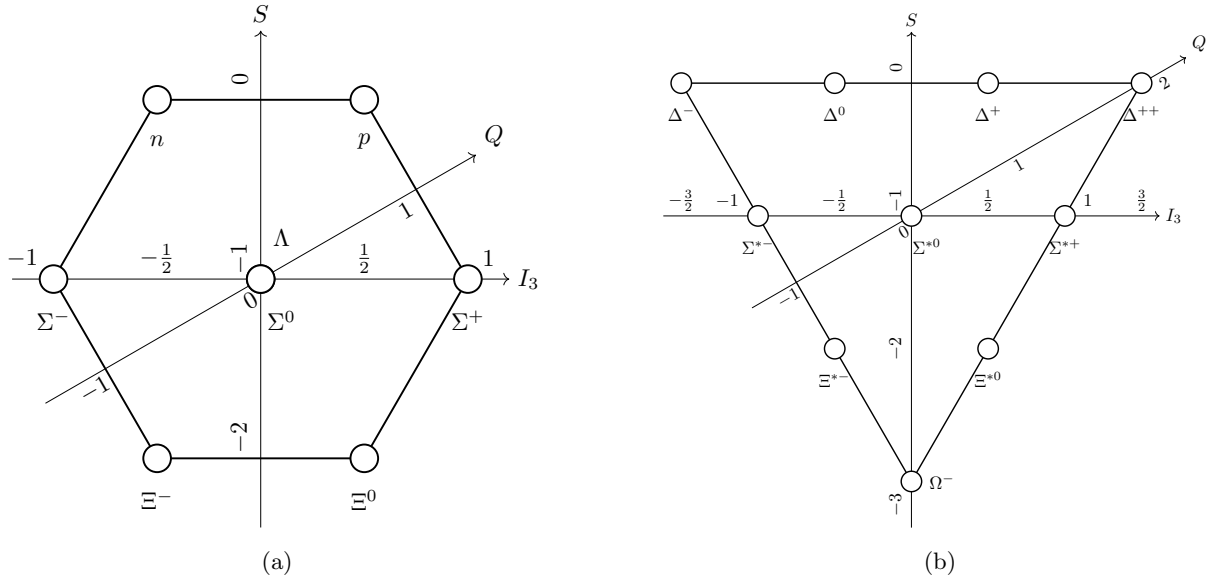
\begin{figure}[tbp]
    \centering
    \subfloat[\label{fig:octet}]{
        \resizebox{0.45\textwidth}{!}{\begin{tikzpicture}[scale=3, every node/.style={font=\small}]
  \draw[->] (0,-1.2) -- (0,1.2) node[above] {\(S\)};
  \draw[->] (-1.2,0) -- (1.2,0) node[right] {\(I_3\)};
  \draw[->] (-1.039,-0.6) -- (1.039,0.6) node[above right] {\(Q\)};
  
  \coordinate (Neutron) at (-0.5,0.866);
  \coordinate (Proton) at (0.5,0.866);
  \coordinate (SigmaMinus) at (-1,0);
  \coordinate (SigmaZero) at (0,0);
  \coordinate (SigmaPlus) at (1,0);
  \coordinate (XiMinus) at (-0.5,-0.866);
  \coordinate (XiZero) at (0.5,-0.866);
  \coordinate (Lambda) at (0,0);

  \draw[thick] (Neutron) -- (SigmaMinus) -- (XiMinus) -- (XiZero) -- (SigmaPlus) -- (Proton) -- (Neutron);

    \foreach \pos/\name in {
      Neutron/$n$,
      Proton/$p$,
      SigmaMinus/$\Sigma^{-}$,
      SigmaPlus/$\Sigma^{+}$,
      XiMinus/$\Xi^{-}$,
      XiZero/$\Xi^{0}$
    } {
      \node[circle, draw, thick, fill=white, minimum size=0.4cm] at (\pos) {};
      \node at ($( \pos ) + (0,-0.2)$) {\name};
    }
    \node[circle, draw, thick, fill=white, minimum size=0.4cm] at (Lambda) {};
    \node at ($(Lambda) + (0.1,0.2)$) {$\Lambda$};
    \node[circle, draw, thick, fill=white, minimum size=0.4cm] at (SigmaZero) {};
    \node at ($(SigmaZero) + (0.1,-0.2)$) {$\Sigma^{0}$};

  \node[rotate=90] at (-0.1,-0.766) {$-2$};
  \node[rotate=90] at (-0.1,0.1) {$-1$};
  \node[rotate=90] at (-0.1,0.966) {0};
  \node at (-1.15,0.1) {$-1$};
  \node at (-0.5,0.1) {$-\frac{1}{2}$};
  \node at (0.5,0.1) {$\frac{1}{2}$};
  \node at (1.1,0.1) {1};
  \node[rotate=30] at (-0.8,-0.533) {$-1$};
  \node[rotate=30] at (-0.05,-0.1) {0};
  \node[rotate=30] at (0.7,0.333) {1};

\end{tikzpicture}}
    }
    \hfill
    \subfloat[\label{fig:decuplet}]{
        \resizebox{0.45\textwidth}{!}{\begin{tikzpicture}[scale=3, every node/.style={font=\small}]
  \draw[->] (0,-2.033) -- (0,1.2) node[above] {\(S\)};
  \draw[->] (-1.617,0) -- (1.617,0) node[right] {\(I_3\)};
  \draw[->] (-1.039,-0.6) -- (1.789,1.033) node[above right] {\(Q\)};
  
  \coordinate (DeltaMinus) at (-1.5,0.866);
  \coordinate (DeltaZero) at (-0.5,0.866);
  \coordinate (DeltaPlus) at (0.5,0.866);
  \coordinate (DeltaPlusPlus) at (1.5,0.866);
  \coordinate (SigmaMinus) at (-1,0);
  \coordinate (SigmaZero) at (0,0);
  \coordinate (SigmaPlus) at (1,0);
  \coordinate (XiMinus) at (-0.5,-0.866);
  \coordinate (XiZero) at (0.5,-0.866);
  \coordinate (OmegaMinus) at (0,-1.732);

  \draw[thick] (DeltaMinus) -- (SigmaMinus) -- (XiMinus) -- (OmegaMinus) -- (XiZero) -- (SigmaPlus) -- (DeltaPlusPlus) -- (DeltaPlus) -- (DeltaZero) -- (DeltaMinus);

    \foreach \pos/\name in {
      DeltaMinus/$\Delta^-$,
      DeltaZero/$\Delta^0$,
      DeltaPlus/$\Delta^+$,
      DeltaPlusPlus/$\Delta^{++}$,
      SigmaMinus/$\Sigma^{*-}$,
      SigmaPlus/$\Sigma^{*+}$,
      XiMinus/$\Xi^{*-}$,
      XiZero/$\Xi^{*0}$
    } {
      \node[circle, draw, thick, fill=white, minimum size=0.4cm] at (\pos) {};
      \node at ($( \pos ) + (0,-0.2)$) {\name};
    }
    \node[circle, draw, thick, fill=white, minimum size=0.4cm] at (SigmaZero) {};
    \node at ($(SigmaZero) + (0.1,-0.2)$) {$\Sigma^{*0}$};
    \node[circle, draw, thick, fill=white, minimum size=0.4cm] at (OmegaMinus) {};
    \node at ($(OmegaMinus) + (0.2,0)$) {$\Omega^{-}$};

  \node[rotate=90] at (-0.1,-1.882) {$-3$};
  \node[rotate=90] at (-0.1,-0.866) {$-2$};
  \node[rotate=90] at (-0.1,0.1) {$-1$};
  \node[rotate=90] at (-0.1,0.966) {0};
  \node at (-1.5,0.1) {$-\frac{3}{2}$};
  \node at (-1.2,0.1) {$-1$};
  \node at (-0.5,0.1) {$-\frac{1}{2}$};
  \node at (0.5,0.1) {$\frac{1}{2}$};
  \node at (1.15,0.1) {1};
  \node at (1.5,0.1) {$\frac{3}{2}$};
  \node[rotate=30] at (-0.8,-0.533) {$-1$};
  \node[rotate=30] at (-0.05,-0.1) {0};
  \node[rotate=30] at (0.7,0.333) {1};
  \node[rotate=30] at (1.65,0.866) {2};

\end{tikzpicture}}
    }
    \caption{Displayed in the plots are (a) the spin-$1/2$ ground-state baryon octet and (b) the spin-$3/2$ baryon decuplet in the SU(3) flavor symmetry framework. The axes, $Q$, $S$, and $I_3$ represent charge, strangeness, and the third component of isospin, respectively.}
    \label{fig:multiplet}
\end{figure}

Demanding the $S$-matrix suppresses spin entanglement leads to emergent symmetries that are not present in the fundamental Lagrangian of QCD. In particular, it was found that~\cite{Low:2021ufv} when the $S$-matrix is an Identity gate in the two-qubit space, the resulting symmetry is the spin-flavor symmetry~\cite{Wigner:1936dx, Wigner:1937zz, Wigner:1939zz, Wagman:2017tmp}, while a SWAP gate gives rise to the non-relativistic conformal invariance~\cite{Birse:1998dk, Mehen:1999nd}. More generally, when the $u, d, s$ quarks are taken to be degenerate in mass, the nucleons are part of the eight-dimensional octet representation under the SU(3) flavor symmetry, which contains other spin-1/2 baryons, as shown in Fig.~\ref{fig:multiplet}(a). Analyzing the 2-to-2 scatterings of the spin-1/2 octet baryons, which include 64 channels in total, revealed more instances of enhanced symmetries due to entanglement suppression in the spin space. This suggests that entanglement suppression may be regarded as an emergent phenomenological property of low-energy hadronic dynamics, realized either when the scattering exhibits larger symmetries than naively expected, or when it approaches the non-relativistic conformal limit. The correlation between entanglement and symmetry goes beyond low-energy QCD, as similar connections were also pointed out in the fully relativistic scattering of two qubits in the context of two-Higgs-doublet models~\cite{Carena:2023vjc,Carena:2025wyh}, which are prototypical models for electroweak symmetry breaking. Moreover, Ref.~\cite{Busoni:2025dns} recently studied the appearance of symmetries in terms of the conservation of non-stabilizerness, also known as magic, during the scattering processes~\cite{Bravyi:2004isx, Robin:2024oqc, White:2024nuc, Chernyshev:2024pqy, Liu:2025frx, Liu:2025qfl, Gargalionis:2025iqs, Liu:2025bgw}.

Building on the observations in the spin-1/2 octet baryons, in this work we will focus on the non-relativistic, $S$-wave scattering of the lowest-lying spin-3/2 light baryons, which form a ten-dimensional representation (decuplet) under the SU(3) flavor symmetry, and study the relation between entanglement suppression and enhanced symmetries. In Fig.~\ref{fig:multiplet}(b) we show the electric charge $(Q)$ and the strangeness quantum number $(S)$ of the decuplet baryons, which are conserved in QCD. Such a system is interesting for a variety of reasons. First, the spin-3/2 baryon can be viewed as a four-dimensional qudit in the spin space\footnote{For brevity we would just refer to spin-3/2 baryons as ``qudits'' henceforth.} and the two-body $S$-matrix would be a two-qudit quantum gate. There have been attempts to study entanglement and symmetry in two-qutrit systems, such as the pion-pion scattering analyzed as a two-qutrit system in the isospin space \cite{Beane:2021zvo}, or the deuteron-deuteron scattering as a system of two spin-1 particles \cite{Kirchner:2023dvg}. However, in the limit of exact isospin symmetry pions are indistinguishable particles under strong interactions. They obey Bose-Einstein (BE) statistics and the $S$-matrix only acts on the sub-Hilbert space with totally symmetric wave functions, which needs to be dealt with care. Similar subtleties arise in identical deuteron-deuteron scattering as well.\footnote{The same considerations apply to the scattering of identical bosons in the two-Higgs-doublet model \cite{Chang:2024wrx}.} In the limit of exact SU(3) flavor symmetry, both the spin-1/2 octet and the spin-3/2 decuplet baryons need to follow the Fermi-Dirac (FD) statistics. The case of spin-1/2 baryons (qubits) has been analyzed in detail in Ref.~\cite{Liu:2022grf}. However, we will see for spin-3/2 baryons (qudits) a rich and intricate interplay among entanglement, quantum statistics and symmetry appears.

Although our calculation is general and could be applied to any two-qudit systems, it is insightful to compare with experimental data, lattice QCD studies and large-$N_c$ analyses on the scattering of spin-3/2 baryons. For instance, based on the deuteron in $^3S_1$ $NN$ scattering and the existence of a virtual state in $^1S_0$ $NN$ scattering, four more zero-strangeness dibaryons were predicted using an SU(6) model long ago~\cite{Dyson:1964xwa}, including two $\Delta\Delta$ dibaryons with $(J,I)=(3,0)$ and $(0,3)$. There has been experimental evidence for the $(J,I)=(3,0)$ dibaryon, namely the $d^{*}(2380)$ reported by the WASA-at-COSY Collaboration~\cite{Bashkanov:2008ih, WASA-at-COSY:2011bjg, WASA-at-COSY:2014dmv}.\footnote{There is also a non-resonant explanation for the experimental signal~\cite{Molina:2021bwp}.} In addition, lattice calculation for the $\Omega\Omega$ system is available for the $^{1}S_{0}$ channel, leading to a large scattering length $a=4.6$~fm~\cite{Gongyo:2017fjb}. Finally, large-$N_{c}$ consequence on the decuplet baryons (mainly in the strangeness $S=0$ sector) has recently been discussed in Ref.~\cite{Richardson:2024zln}.

The paper is organized as follows. In Section \ref{sec:S}, we present the general construction of the $S$-matrix for the non-relativistic scattering of spin-3/2 decuplet baryons. In Section \ref{sec:EP}, the definition and detailed calculation of the entanglement power in the spin space for distinguishable particles are discussed. In Section \ref{sect:4}, we discuss the conditions for entanglement suppression in decuplet-baryon scatterings, followed by Section \ref{sect:group} for the identification of enhanced symmetries using group-theoretic reasoning. These symmetries are verified in the Lagrangian approach using non-relativistic effective field theories in Section \ref{sec:EFT}. Then in Section \ref{sec:subS} we present a dedicated analysis for identical particle scatterings. Summary and Outlook are given in Section \ref{sec:sum}. We also present two appendices: Appendix \ref{app1} on the classification of the spin and flavor quantum numbers for decuplet scatterings and Appendix \ref{app2} on the explicit calculation of entanglement power in neutron-neutron scattering.

\section{$S$-matrix for $S$-wave Scatterings}\label{sec:S}

In this paper, we primarily study the scattering of two light decuplet baryons in the low-energy regime,\footnote{In principle, the octet-octet and octet-decuplet channels with lower thresholds can also couple to the decuplet-decuplet systems with the same quantum numbers. In the near-threshold region of the decuplet pair, such channels do not produce any sharp energy dependence, and thus effectively give an optical potential to the decuplet-decuplet interaction, manifesting as an imaginary part of the contact terms. For simplicity, we disregard the complexities associated with this imaginary component.} which is dominated by $S$-wave interactions. Therefore the total angular momentum of the system is given by the total spin of the two baryons. Also, we consider the leading order (LO) of a non-relativistic effective field theory (NREFT). At this order, all decuplet baryons are stable as their widths come from self-energy corrections at loop level, and no SU(3) breaking operators exist as such operators must contain powers of the quark mass matrix and are subleading, suppressed by $\mathcal{O}(m_s/\Lambda_{\rm QCD})$ with $m_s$ and $\Lambda_{\rm QCD}$ being the strange quark mass and the non-perturbative QCD scale, respectively. Thus, we proceed with the discussion assuming exact flavor SU(3) symmetry though the effects of symmetry breaking can be important in certain aspects of actual baryon scatterings. Consequently, all emergent symmetries predicted in this framework should also be regarded as approximate in reality, just as SU(3) flavor symmetry itself is.

Our focus lies in understanding the spin and flavor symmetries of the system, so we define our $S$-matrix as
\begin{equation}\label{eq:Sdef}
\hat S=\sum_{J,F}{\mathcal{J}_{J}\otimes\mathcal{F}_{F}}\,e^{2i\delta _{JF}},
\end{equation}
where $\mathcal{J}_{J}\otimes\mathcal{F}_{F}$'s denote the projection operators onto subspaces belonging to definite irreducible representations (irreps) of the spin group SU(2) and the flavor group SU(3). Specifically, $J$ can be regarded as the total spin of this irrep. Therefore, the $S$-matrix is constructed using these projectors and the corresponding phase shifts $\delta _{JF}(p)$'s as in Eq.~\eqref{eq:Sdef}, which in general depend on the center-of-mass momentum $p$ of the scattering, although we often suppress the $p$-dependence for brevity.

The decomposition of $S$-matrix into irreps is determined from group theory, by the tensor product of the spin and flavor properties of the individual particles in the 2-to-2 scattering, similar to the addition of angular momenta in quantum mechanics. For two particles belonging to the irreps $A$ and $B$, the tensor product $A\otimes B$ can be decomposed into a direct sum of irreps:
\begin{equation}
A\otimes B=\oplus_iC_i\,.
\end{equation}
The operator $\mathcal P_{C_i}$ projects the tensor product $A\otimes B$ into a specific irrep $C_i$:
\begin{equation}
\mathcal P_{C_i}|C_j\rangle=\delta_{ij}|C_j\rangle\,,\quad \mathcal P_{C_i}|A\rangle\otimes|B\rangle=|C_i\rangle\,.
\end{equation}
The projectors can be expressed as polynomials of Casimir operators $\boldsymbol{a}\cdot\boldsymbol{b}\equiv \sum_\alpha a^\alpha \otimes b^\alpha$ whose eigenvalues label the irreps. Here $\boldsymbol{a} = \left\{a^\alpha\right\}$ and $\boldsymbol{b}=\left\{b^\alpha\right\}$ denote the generators of the spin SU(2) or flavor SU(3) in the corresponding representations $A$ and $B$.

Focusing on the spin space for demonstration purposes, let us consider the scattering of two distinguishable spin-1/2 particles, such as the neutron-proton scattering in the $S$-wave, where the total angular momentum is simply given by the total spin:
\begin{equation}
\frac 1 2 \otimes \frac 1 2=0 \oplus 1\,,
\end{equation}
so there are two projectors for the two total spin values $J=0$ and 1. The Casimir operator is $\boldsymbol{t}_{1/2}\cdot\boldsymbol{t}_{1/2}$ with $\boldsymbol{t}_{s}$ the SU(2) generators in the $(2s+1)$-dimensional representation. Therefore, $\boldsymbol{t}_{1/2}=\boldsymbol{\sigma }/2$ with $\boldsymbol{\sigma }$ the Pauli matrices. Now we can write down the $S$-matrix for a $1/2\otimes 1/2$ system:
\begin{equation}\label{eq:S1/2}
\hat S=\mathcal{J}_{0}\,e^{2i\delta _{0}}+\mathcal{J}_{1}\,e^{2i\delta _{1}},
\end{equation}
where
\begin{equation}
\mathcal{J}_{0}=\frac{1}{4}\left[ 1-4\left(\boldsymbol{t}_{1/2}\cdot \boldsymbol{t}_{1/2}\right) \right] ,\quad \mathcal{J}_{1}=\frac{1}{4}\left[ 3+4\left(\boldsymbol{t}_{1/2}\cdot \boldsymbol{t}_{1/2}\right) \right]. 
\end{equation}
It is easy to check
\begin{align}
&\mathcal{J}_{0}|J=0\rangle=|J=0\rangle\,,&&\mathcal{J}_{0}|J=1\rangle=0\,,\\
&\mathcal{J}_{1}|J=0\rangle=0\,,&&\mathcal{J}_{1}|J=1\rangle=|J=1\rangle\,,
\end{align}
where $|J=0\rangle$ and $|J=1\rangle$ denote the states with total spin $J=0$ and $J=1$, respectively.

The discussion can be straightforwardly extended to two distinguishable spin-3/2 particle scatterings in the $S$-wave. The total spin of two such particles ranges from $0$ to $3$, which indicates that the projectors should be cubic polynomials of the Casimir operator. They read as
\begin{align}\label{Eq:spinproj}
\mathcal{J}_{0}&=\frac{33}{128} +\frac{31}{96}\left(\boldsymbol{t}_{3/2}\cdot \boldsymbol{t}_{3/2}\right)-\frac{5}{72}\left(\boldsymbol{t}_{3/2}\cdot \boldsymbol{t}_{3/2}\right)^2-\frac{1}{18}\left(\boldsymbol{t}_{3/2}\cdot \boldsymbol{t}_{3/2}\right)^3  ,\\
\mathcal{J}_{1}&=-\frac{81}{128} -\frac{117}{160}\left(\boldsymbol{t}_{3/2}\cdot \boldsymbol{t}_{3/2}\right)+\frac{9}{40}\left(\boldsymbol{t}_{3/2}\cdot \boldsymbol{t}_{3/2}\right)^2+\frac{1}{10}\left(\boldsymbol{t}_{3/2}\cdot \boldsymbol{t}_{3/2}\right)^3  ,\\
\mathcal{J}_{2}&=\frac{165}{128}+\frac{23}{96}\left(\boldsymbol{t}_{3/2}\cdot \boldsymbol{t}_{3/2}\right)-\frac{17}{72}\left(\boldsymbol{t}_{3/2}\cdot \boldsymbol{t}_{3/2}\right)^2-\frac{1}{18}\left(\boldsymbol{t}_{3/2}\cdot \boldsymbol{t}_{3/2}\right)^3  ,\\
\mathcal{J}_{3}&=\frac{11}{128}+\frac{27}{160}\left(\boldsymbol{t}_{3/2}\cdot \boldsymbol{t}_{3/2}\right)+\frac{29}{360}\left(\boldsymbol{t}_{3/2}\cdot \boldsymbol{t}_{3/2}\right)^2+\frac{1}{90}\left(\boldsymbol{t}_{3/2}\cdot \boldsymbol{t}_{3/2}\right)^3,
\end{align}
where $\boldsymbol{t}_{3/2}$ represent the SU(2) generators in the four-dimensional representation. The $S$-matrix in the spin space is then
\begin{equation}\label{eq:S3/2}
\hat S=\mathcal{J}_{0}\,e^{2i\delta _{0}}+\mathcal{J}_{1}\,e^{2i\delta _{1}}+\mathcal{J}_{2}\,e^{2i\delta _{2}}+\mathcal{J}_{3}\,e^{2i\delta _{3}}.
\end{equation}

Of course, the full decuplet-decuplet $S$-matrix involves both SU(2) (spin) and SU(3) (flavor) projectors. However, as we will demonstrate in Section~\ref{sec:theo}, when a system is constrained by quantum statistics, as is the case for octet and decuplet baryons in the limit of exact SU(3) flavor symmetry, it is often sufficient to consider only the spin space (or, equivalently, only the flavor space), which greatly simplifies the analysis.

\section{Entanglement Power for Distinguishable Particles}\label{sec:EP}

Entanglement measures have been developed to quantify the degree of entanglement present in a given quantum system. In this section, we focus on systems with distinguishable particles, and identical particles will be discussed in Section~\ref{sec:theo}.

For a bipartite state $| \psi \rangle \in {\cal H}_A\otimes {\cal H}_B$, a widely used entanglement measure is the linear entropy, defined as (see, e.g., Refs.~\cite{Beane:2021zvo, Liu:2022grf})
\begin{equation}
\mathcal{E}(| \psi \rangle) = 1 - \operatorname{Tr}_A\left[\rho_A^2\right],
\end{equation}
where $\rho=| \psi \rangle \left< \psi \right|$ is the density matrix, and $\rho_A = \operatorname{Tr}_B[\rho]$ is the reduced density matrix obtained after tracing over subsystem $B$. $\mathcal E(| \psi \rangle)$ is semi-positive definite and vanishes only when $| \psi \rangle$ is a tensor product, i.e., $| \psi \rangle=| \psi_A \rangle\otimes| \psi_B \rangle$, as shown in the appendix of Ref.~\cite{Hu:2024hex}.

While entanglement measure describes how entangled a given state is, entanglement power characterizes the capacity of an operator $\hat U$ to generate entanglement. It is defined by averaging the entanglement of the output states produced by acting $\hat U$ on all possible product states~\cite{Zanardi:2001zza}:
\begin{equation}
\label{eq:euave}
E(\hat U) = \overline{\mathcal{E}\left( \hat U| \psi_A \rangle\otimes| \psi_B \rangle\right) } \equiv  \int\mathrm{d}\omega_A\mathrm{d}\omega_B\,\mathcal{E}\left( \hat U| \psi_A \rangle\otimes| \psi_B \rangle\right),
\end{equation}
where the integral is taken over all elements belonging to irreps $A$ and $B$, with $\mathrm{d}\omega_{A,B}$ denoting the corresponding normalized integration measures: 
\begin{equation}
\int \mathrm{d}\omega_{A}= \int \mathrm{d}\omega_{B}= 1\,.
\end{equation}
In general, a quantum state in an $(m+1)$-dimensional Hilbert space resides on the manifold $\mathbb{CP}^{m}$, which can be parameterized as:
\begin{equation}
\label{eq:mpsi}
| \psi_{\bm{m+1}} \rangle =\left(n_0,n_1e^{i\nu_1},\ldots,n_me^{i\nu_m}\right),\quad\nu_i\in[0,2\pi)\,,
\end{equation}
where
\begin{equation}
\left. \left\{ \begin{matrix}
	n_0&		=&		\cos \theta _1\sin \theta _2\sin \theta _3\ldots\sin \theta _m\\
	n_1&		=&		\sin \theta _1\sin \theta _2\sin \theta _3\ldots\sin \theta _m\\
	n_2&		=&		\cos \theta _2\sin \theta _3\ldots\sin \theta _m\\
	\cdots&		&		\cdots\\
	n_m&		=&		\cos \theta _m\\
\end{matrix} \right. \right.,\quad\theta_i\in
\left[ 0,\frac{\pi}{2} \right).
\end{equation}
The commonly used Fubini-Study measure is~\cite{Bengtsson:2001yd, Bengtsson:2006rfv}
\begin{equation}
\mathrm{d}\omega_{\bm{m+1}}=\frac{m!}{\pi^m}\prod_{i=1}^{m} \,\mathrm{d}\theta_{i} \,\mathrm{d}\nu_{i}\,\cos\theta_{i}\sin^{2i-1}\theta_{i}\,, \quad \int \mathrm{d}\omega_{\bm{m+1}} = 1\,,
\end{equation}
which is properly normalized. Note that for a spin-$s$ particle, $m=2s$.

Entanglement power serves as a state-independent measure of entanglement generation, which is semi-positive definite and vanishes if and only if $\hat U| \psi \rangle$ remains a tensor-product state for all $| \psi \rangle=| \psi_A \rangle\otimes| \psi_B \rangle$. The entanglement power $E(\hat S)$ of an $S$-matrix can then be defined as
\begin{equation}\label{eq:EPdef}
E(\hat S)=1-\int\mathrm{d}\omega_A\mathrm{d}\omega_B\operatorname{Tr}_A\left[\rho_A^2\right],
\end{equation}
where
\begin{equation}\label{eq:rho}
\rho =|\psi _{\mathrm{out}}\rangle \langle \psi _{\mathrm{out}}|\,,\quad|\psi _{\mathrm{out}}\rangle =\hat S|\psi _{\mathrm{in}}\rangle\,,\quad|\psi _{\mathrm{in}}\rangle =|\psi _A\rangle \otimes |\psi _B\rangle\,.
\end{equation}
For distinguishable particles the $S$-matrix acts on the full Hilbert space spanned by the initial state $|\psi_{\mathrm{in}}\rangle$. The unitarity of the $S$-matrix ensures that the density matrix is properly normalized: ${\rm Tr}[\rho] =1$. In the spin space, this averaging integration in Eq.~\eqref{eq:EPdef} can be understood as originating from the fact that the directions of the spins in the initial states can be arbitrary.

We first take the $1/2\otimes 1/2$ system for example. $|\psi _A\rangle$ and $|\psi _B\rangle $ are two distinguishable spin-1/2 particles. Setting $m+1=2s+1=2$ in Eq.~\eqref{eq:mpsi},
\begin{equation}\label{eq:1/2rep}
	| \psi_{\bm{2}} \rangle= \left(\cos \theta_1, \sin \theta_1\, e^{i\nu_1}\right)^T,
\end{equation}
with $\theta_1\in[0,\pi/2)$ and $\nu_1\in[0,2\pi)$. The integral measure $\mathrm{d}\omega_{\bm{2}}$ then reads as
\begin{equation}
    \mathrm{d}\omega_{\bm{2}} = \frac1\pi \mathrm{d}\theta_1 \cos\theta_1\sin\theta_1 \mathrm{d}\nu_1 \,.
\end{equation}
This parametrization is identical to the usual Bloch sphere parametrization for a qubit with $(\theta,\phi)$ as $\theta_1=\theta/2$ and $\nu_1=\phi$.

Now we can calculate the entanglement power for various cases. For distinguishable particles, the result for the $1/2\otimes1/2$ $S$-matrix in Eq.~\eqref{eq:S1/2} has been known~\cite{Beane:2018oxh, Liu:2022grf}:
\begin{equation}
E(\hat S)=\frac{1}{6}\sin^2\left[2\left(\delta _0-\delta _1\right)\right],
\end{equation}
and $E(\hat S)=0$ yields two possible solutions (modulo multiple times $\pi$):
\begin{equation}\label{eq:1/2delta}
|\delta_0-\delta_1|=0\,,\quad\text{or}\quad\frac{\pi}{2}\,.
\end{equation}
Note that we restrict the phase shifts to the range $\delta\in[0,\pi)$ because they only appear in the form of $e^{2i\delta}$. Moreover, the solutions for $|\delta_0-\delta_1|=\pi/2$ are $\delta_0=0$ and $\delta_1=\pi/2$ (or vice versa), as will be proven in Sec.~\ref{sec:EFT}. We highlight here these two particularly important phase shift values associated with the non-interacting limit and unitary limit\footnote{At the unitarity limit, $p\cot\delta=0$, where $p$ is the magnitude of the center-of-mass momentum and $\delta$ is the phase shift, so that the inverse of the scattering amplitude is given only the unitarity term, and correspondingly $\delta=\pi/2$.}, respectively; see more discussions in Sec.~\ref{sec:EFT}.

For the scattering of distinguishable spin-3/2 particles, the $S$-matrix is given in Eq.~\eqref{eq:S3/2} and we obtain
\begin{align}
E(\hat S)&=\frac{1}{200000}\Big\{ 77482-2100\cos \left[ 2\left( \delta _0+\delta _1-\delta _2-\delta _3 \right) \right]\notag\\
&\quad -2100\cos \left[ 2\left( \delta _0-\delta _1+\delta _2-\delta _3 \right)\right]-2100\cos \left[ 2\left( \delta _0-\delta _1-\delta _2+\delta _3 \right) \right]\notag\\
&\quad -1200\cos \left[ 2\left( \delta _0-2\delta _1+\delta _2 \right) \right]-4200\cos \left[ 2\left( \delta _0+\delta _2-2\delta _3 \right) \right]\notag\\
&\quad -8400\cos \left[ 2\left( \delta _1-2\delta _2+\delta _3 \right) \right]\notag\\
&\quad -375\cos \left[ 4\left( \delta _0-\delta _1 \right) \right]-10800\cos \left[ 2\left( \delta _0-\delta _2 \right) \right] -625\cos \left[ 4\left( \delta _0-\delta _2 \right) \right] \notag\\
&\quad -875\cos \left[ 4\left( \delta _0-\delta _3 \right) \right]-2175\cos \left[ 4\left( \delta _1-\delta _2 \right) \right]-26376\cos \left[ 2\left( \delta _1-\delta _3 \right) \right]\notag\\
&\quad -5481\cos \left[ 4\left( \delta _1-\delta _3 \right) \right]-10675\cos \left[ 2\left( \delta _2-\delta _3 \right) \right]\Big\},
\end{align}
which vanishes when all the cosines are valued at +1 and leads to the following non-entangling solutions:
\begin{equation}\label{eq:3/2delta}
\delta_0=\delta_2=\delta_\text{even}\,,\qquad\delta_1=\delta_3=\delta_\text{odd}\,;\qquad|\delta_\text{even}-\delta_\text{odd}|=0\,,\quad\text{or}\quad\frac{\pi}{2}\,.
\end{equation}

We observe an intriguing feature of the solutions: the phase shifts for all channels with even total spin are equal, and similarly, those for odd total spin are also equal, with the two sets differing by either 0 or $\pi/2$. This turns out to be a general pattern which we prove in the next section.

\section{Entanglement Suppression in Decuplet Scattering}\label{sect:4}

\subsection{A General Result for Two Qudits}\label{sec:theo}

In this section, we begin with the observation made at the end of the previous section and formulate a general result concerning entanglement suppression for two-qudit systems. Although we focus on spin-3/2 particles as four-dimensional qudits, the generalization to arbitrary qudits is straightforward. Additionally, we  consider the connection between quantum statistics  and quantum entanglement. The discussion here provides an alternative perspective to the general reasoning in Ref.~\cite{McGinnis:2025brt}.

A natural starting point is to consider the form of the operators implied by the phase-shift relations in Eq.~\eqref{eq:3/2delta}. If $|\delta_\text{even}-\delta_\text{odd}|=0$, meaning all four phase shifts are equal, the resulting $S$-matrix is simply an Identity operator with an overall phase, which clearly has zero entanglement power (it leaves any tensor-product state invariant). The other solution $|\delta_\text{even}-\delta_\text{odd}|=\pi/2$ is particularly interesting, with the non-entangling $S$-matrix proportional to
\begin{equation}\label{eq:SWAP3o2}
-\left(\mathcal{J}_{0}-\mathcal{J}_{1}+\mathcal{J}_{2}-\mathcal{J}_{3}\right)=\sum_{\text{odd}~J}\mathcal{J}_{J}-\sum_{\text{even}~J}\mathcal{J}_{J}\,,
\end{equation}
Note that channels with an odd total spin $J$ correspond to the symmetric part of the $3/2\otimes3/2$ tensor-product space, while those with an even $J$ correspond to the antisymmetric part, so the above operator is precisely a SWAP gate~\cite{Low:2021ufv}, which is a quantum logic gate exchanging the two substates, thus yielding 1 when acting on symmetric states and $-1$ on antisymmetric states.

In general, the SWAP operator can be defined as follows for a bipartite system with equal dimensions: ${\cal H}_R\otimes {\cal H}_R$, where $R$ is a certain irrep of a group $G$. The tensor product decomposes as
\begin{equation}
R\otimes R=\left(\oplus_iS_i\right)\oplus\left(\oplus_jA_j\right),
\end{equation}
where $S_i$ and $A_j$ denote symmetric and antisymmetric irreps, respectively. The SWAP is then given by the sum of projectors onto symmetric irreps ($\mathcal{S}_{i}$) minus the sum of projectors onto antisymmetric ones ($\mathcal{A}_{j}$):
\begin{equation}\label{eq:SWAPdef}
\operatorname{SWAP}=\sum_{i}{\mathcal{S}_{i}}-\sum_{j}{\mathcal{A}_{j}} \equiv {\cal P}_{\rm S} - {\cal P}_{\rm A}\,; \quad \sum_{i}{\mathcal{S}_{i}}\equiv {\cal P}_{\rm S}\,, \quad \sum_{j}{\mathcal{A}_{j}} \equiv {\cal P}_{\rm A}\,,
 \end{equation}
where ${\cal P}_{\rm S}$ and ${\cal P}_{\rm A}$ are the projection operators into the symmetric and  antisymmetric part of the bipartite Hilbert space, respectively,
\begin{equation}
    {\cal P}_{\rm S}\, |S_i\rangle = |S_i\rangle \,, \quad  {\cal P}_{\rm S}\, |A_i\rangle =0\,;\qquad {\cal P}_{\rm A} |A_i\rangle = |A_i\rangle \,, \quad {\cal P}_{\rm A} |S_i\rangle= 0  \,,
\end{equation}
which implies that $|S_i\rangle$ and $|A_j\rangle$ are eigenstates of SWAP with $+1$ and $-1$ eigenvalues:
\begin{equation}
\label{eq:swapdef}
\operatorname{SWAP}|S_i\rangle=|S_i\rangle\,,\quad\operatorname{SWAP}|A_j\rangle=-|A_j\rangle\,.
\end{equation}
Note ${\cal P}_{\rm S}$ and ${\cal P}_{\rm A}$ are complete in the sense that ${\cal P}_{\rm S}+{\cal P}_{\rm A}=1$ is the Identity operator over the entire Hilbert space. In fact, if we choose a basis consisting of symmetric and antisymmetric base vectors, ${\cal P}_{\rm S}$ is an Identity operator over the symmetric part of the Hilbert space and vanishes over the antisymmetric part. Conversely for ${\cal P}_{\rm A}$.

Under the definition in Eq.~\eqref{eq:SWAPdef}, one can show that SWAP interchanges the two qudits.
Let us consider an arbitrary initial product state written as
\begin{equation}
|\psi\rangle=|a\rangle\otimes |b\rangle\,, \quad |a\rangle=\sum_i{a_i|\phi_i\rangle}\,, \quad|b\rangle= \sum_j{b_j|\phi_j\rangle}\,.
\end{equation}
where $\{|\phi_i\rangle\}$ is a set of basis vectors for ${\cal H}_R$.
The SWAP gate acts on it as follows:
\begin{align}
\operatorname{SWAP}|\psi\rangle&=\operatorname{SWAP}\left(\sum_i{a_i|\phi_i\rangle}\otimes \sum_j{b_j|\phi_j\rangle}\right)\notag\\
&=\operatorname{SWAP}\left(\sum_{i,j}{\frac{a_ib_j+b_ia_j}{2}|\phi_i\rangle \otimes |\phi_j\rangle}+\sum_{i,j}{\frac{a_ib_j-b_ia_j}{2}|\phi_i\rangle \otimes |\phi_j\rangle}\right)\notag\\
&=\sum_{i,j}{\frac{a_ib_j+b_ia_j}{2}|\phi_i\rangle \otimes |\phi_j\rangle}-\sum_{i,j}{\frac{a_ib_j-b_ia_j}{2}|\phi_i\rangle \otimes |\phi_j\rangle}\notag\\
&=|b\rangle\otimes |a\rangle\,,
\end{align}
which is simply the product state with the two qudits interchanged.

It has also been shown that, for bipartite systems ${\cal H}_R\otimes {\cal H}_R$, the Identity and the SWAP are the only two gates that would always map a product state to another product state \cite{10.1063/1.3399808, Johnston01102011, 10.1063/1.3578015, Low:2021ufv}; these are the only two minimally entangling quantum gates. If we consider the group $G={\rm SU(2)}_{\rm spin}$, this directly provides the minimal-entanglement condition for the $S$-wave scattering between two spin-$s$ particles. The total spin can range from 0 to $2s$, corresponding to $(2s+1)$ phase shifts. Entanglement power of its $S$-matrix vanishes, $E(\hat{S})=0$, if and only if the phase shifts for even total spins are all equal ($\delta_\text{even}$), the phase shifts for odd total spins are all equal ($\delta_\text{odd}$), and
\begin{equation}\label{eq:sdelta}
|\delta_\text{even}-\delta_\text{odd}|=0\,,\quad\text{or}\quad\frac{\pi}{2}\,.
\end{equation}
This is the entanglement-suppression condition for distinguishable particles in the spin space.

In the cases of octet-octet and decuplet-decuplet scatterings, the $S$-matrix involves the projection operators in both the spin and flavor spaces, as shown in Eq.~\eqref{eq:Sdef}. In this case, the group $G={\rm SU(2)}_{\rm spin}\otimes {\rm SU(3)}_{\rm flavor}$ and an irrep $R$ of $G$ can be written as the tensor product of irreps $R=R_{\rm spin}\otimes R_{\rm flavor}$. As we focus on minimal entanglement in the spin space, it is convenient to define the Identity and SWAP operators acting on the spin and the flavor spaces separately. For instance, consider the action of $\operatorname{SWAP}_\text{spin}\otimes\operatorname{SWAP}_\text{flavor}$ on the qudits:
 \begin{equation}
\operatorname{SWAP}_\text{spin}\otimes\operatorname{SWAP}_\text{flavor}|s_1,f_1\rangle\otimes|s_2,f_2\rangle=|s_2,f_2\rangle\otimes|s_1,f_1\rangle\,,
\end{equation}
where $s_i$ and $f_i$ refer to the spin and flavor quantum numbers, respectively. 

If the qudits further obey quantum statistics---BE for bosons or FD for fermions---we obtain
\begin{equation}
|s_2,f_2\rangle\otimes |s_1,f_1\rangle=
\begin{cases}
+|s_1,f_1\rangle\otimes |s_2,f_2\rangle\,,\quad\text{BE}\,,\\
-|s_1,f_1\rangle\otimes |s_2,f_2\rangle\,,\quad\text{FD}\,.\\
\end{cases}
\end{equation}
Therefore, we immediately conclude~\cite{Liu:2022grf}
\begin{equation}
\label{eq:swap2}
\operatorname{SWAP}_\text{spin}\otimes\operatorname{SWAP}_\text{flavor}=\begin{cases}
+1_{\rm spin}\otimes1_{\rm flavor}\,,\quad\text{BE}\,,\\
-1_{\rm spin}\otimes1_{\rm flavor}\,,\quad\text{FD}\,.\\
\end{cases}
\end{equation}
Moreover, one can similarly derive
\begin{equation}
\operatorname{SWAP}_\text{spin}\otimes1_{\rm flavor}=\begin{cases}
+1_{\rm spin}\otimes\operatorname{SWAP}_\text{flavor}\,,\quad\text{BE}\,,\\
-1_{\rm spin}\otimes\operatorname{SWAP}_\text{flavor}\,,\quad\text{FD}\,.
\end{cases}
\end{equation}
From the above discussion, we arrive at the following conclusion: in the presence of quantum statistics, the ability of an operator to generate entanglement in the spin space is equivalent to its ability to entangle in the flavor space~\cite{Liu:2022grf}, or more generally in the rest of all other spaces. This will greatly simplify the analysis of decuplet-baryon scattering in the next section.

\subsection{Spin-3/2 Baryon Scatterings}\label{sec:ES}

Light decuplet baryons have spin 3/2 and belong to the $\bm{10}$ irrep of the SU(3) flavor group. The spin and flavor structures of the decuplet-decuplet system can be decomposed respectively as
\begin{align}
\frac32\otimes\frac32&=\underbrace{0\oplus2}_{\text{antisymmetric}}\oplus\underbrace{1\oplus3}_{\text{symmetric}},\\
\bm{10}\otimes\bm{10}&=\underbrace{\bm{27}\oplus\bm{28}}_{\text{symmetric}}\oplus\underbrace{\overline{\bm{10}}\oplus\bm{35}}_{\text{antisymmetric}}.
\end{align}
Due to FD statistics, spin-symmetric components project into flavor-antisymmetric irreps, while spin-antisymmetric components project into flavor-symmetric irreps. Hence, the corresponding decuplet-decuplet $S$-matrix takes the form:
\begin{align}\label{eq:Sdd}
\hat S&=\mathcal{J}_{0}\otimes
\left(\mathcal{F}_{\bm{27}}\,e^{2i\delta_{0,\bm{27}}}+\mathcal{F}_{\bm{28}}\,e^{2i\delta_{0,\bm{28}}}\right) \notag\\
&\quad +\mathcal{J}_{1}\otimes
\left(\mathcal{F}_{\overline{\bm{10}}}\,e^{2i\delta_{1,\overline{\bm{10}}}}+\mathcal{F}_{\bm{35}}\,e^{2i\delta_{1,\bm{35}}}\right) \notag\\
&\quad +\mathcal{J}_{2}\otimes
\left(\mathcal{F}_{\bm{27}}\,e^{2i\delta_{2,\bm{27}}}+\mathcal{F}_{\bm{28}}\,e^{2i\delta_{2,\bm{28}}}\right) \notag\\
&\quad +\mathcal{J}_{3}\otimes
\left(\mathcal{F}_{\overline{\bm{10}}}\,e^{2i\delta_{3,\overline{\bm{10}}}}+\mathcal{F}_{\bm{35}}\,e^{2i\delta_{3,\bm{35}}}\right) ,
\end{align}
where ${\cal F}_F$ is the projector into a certain flavor irrep $F$. Recall that $\sum_J {\cal J}_J =1_{\rm spin}$ and $\sum_F {\cal F}_F =1_{\rm flavor}$.

As discussed in the previous section, the non-entangling $S$-matrix in the spin space must be proportional to the Identity or SWAP gate in the spin space. The Identity gate can be achieved when all eight phase shifts are equal:
\begin{equation}\label{eq:deltasol1}
\delta_{0,\bm{27}}=\delta_{2,\bm{27}}=\delta_{0,\bm{28}}=\delta_{2,\bm{28}}=\delta_{1,\overline{\bm{10}}}=\delta_{3,\overline{\bm{10}}}=\delta_{1,\bm{35}}=\delta_{3,\bm{35}}\, ,
\end{equation}
in which case
\begin{align}
\hat S&\propto\left(\mathcal{J}_{0}+\mathcal{J}_{2}\right)\otimes\left(\mathcal{F}_{\bm{27}}+\mathcal{F}_{\bm{28}}\right)+\left(\mathcal{J}_{1}+\mathcal{J}_{3}\right)\otimes\left(\mathcal{F}_{\overline{\bm{10}}}+\mathcal{F}_{\bm{35}}\right)\notag\\
&=\frac{1_{\rm spin}-\operatorname{SWAP}_\text{spin}}{2}\otimes\frac{1_{\rm flavor}+\operatorname{SWAP}_\text{flavor}}{2}+\frac{1_{\rm spin}+\operatorname{SWAP}_\text{spin}}{2}\otimes\frac{1_{\rm flavor}-\operatorname{SWAP}_\text{flavor}}{2}\notag\\
&=\frac{1}{2}\left[1_{\rm spin}\otimes1_{\rm flavor}-\operatorname{SWAP}_\text{spin}\otimes\operatorname{SWAP}_\text{flavor}\right]\notag\\
&=1_{\rm spin}\otimes1_{\rm flavor} \,,
\end{align}
where in the second line above we have used the definition of a SWAP gate in Eq.~\eqref{eq:SWAPdef} to write $\operatorname{SWAP}_\text{spin}$ and $\operatorname{SWAP}_\text{flavor}$  as the difference between the projectors in the symmetric and the antisymmetric irreps,
\begin{equation}
\operatorname{SWAP}_\text{spin}=-\left(\mathcal{J}_{0}-\mathcal{J}_{1}+\mathcal{J}_{2}-\mathcal{J}_{3}\right)\,,\quad\operatorname{SWAP}_\text{flavor}=\mathcal{F}_{\bm{27}}+\mathcal{F}_{\bm{28}}-\mathcal{F}_{\overline{\bm{10}}}-\mathcal{F}_{\bm{35}} \,,
\end{equation}
while in the last line we have imposed FD statistics on the decuplet baryons and applied Eq.~\eqref{eq:swap2}. On the other hand, the SWAP$_{\rm spin}$ gate can be achieved only when
\begin{align}\label{eq:deltasol2}
&\delta_{0,\bm{27}}=\delta_{2,\bm{27}}=\delta_{0,\bm{28}}=\delta_{2,\bm{28}}\,,\notag\\
&\delta_{1,\overline{\bm{10}}}=\delta_{3,\overline{\bm{10}}}=\delta_{1,\bm{35}}=\delta_{3,\bm{35}}\,,\notag\\
&\left|\delta_{0,\bm{27}}-\delta_{1,\overline{\bm{10}}}\right|=\frac\pi2\,,
\end{align}
so that
\begin{align}
\hat S&\propto-\left(\mathcal{J}_{0}+\mathcal{J}_{2}\right)\otimes\left(\mathcal{F}_{\bm{27}}+\mathcal{F}_{\bm{28}}\right)+\left(\mathcal{J}_{1}+\mathcal{J}_{3}\right)\otimes\left(\mathcal{F}_{\overline{\bm{10}}}+\mathcal{F}_{\bm{35}}\right)\notag\\
&=-\frac{1_{\rm spin}-\operatorname{SWAP}_\text{spin}}{2}\otimes\frac{1_\text{flavor}+\operatorname{SWAP}_\text{flavor}}{2}+\frac{1_{\rm spin}+\operatorname{SWAP}_\text{spin}}{2}\otimes\frac{1_\text{flavor}-\operatorname{SWAP}_\text{flavor}}{2}\notag\\
&=\frac{1}{2}\left[\operatorname{SWAP}_\text{spin}\otimes1_\text{flavor}-1_\text{spin}\otimes\operatorname{SWAP}_\text{flavor}\right]\notag\\
&=\operatorname{SWAP}_\text{spin}\otimes1_\text{flavor}\,.
\end{align}

Although we include in Eq.~\eqref{eq:Sdd} all possible flavor irreps, not all scattering channels involve every irrep.  This is because scattering processes in QCD conserve both electric charge $Q$ and strangeness $S$. As a result, an initial baryon pair in a given $(Q,S)$ sector can only scatter into final states within the same sector. Consequently, imposing vanishing entanglement power on these channels yields less stringent constraints than in Eqs.~\eqref{eq:deltasol1} and~\eqref{eq:deltasol2}, a fact that has been observed in the scattering of spin-1/2 octet baryons \cite{Liu:2022grf}. For instance, in the $(Q,S)=(0,0)$ sector, which includes two pairs $\left(\Delta^-\Delta^+, \Delta^0\Delta^0\right)$, only $\bm{27}$, $\bm{28}$ and $\bm{35}$ irreps are involved. Therefore, in this sector, the condition of entanglement suppression leads to:
\begin{equation}
\delta_{0,\bm{27}}=\delta_{2,\bm{27}}=\delta_{0,\bm{28}}=\delta_{2,\bm{28}}\,,\qquad\delta_{1,\bm{35}}=\delta_{3,\bm{35}}\,;\qquad|\delta_{0,\bm{27}}-\delta_{1,\bm{35}}|=0\,,\quad\text{or}\quad\frac\pi2\,.
\end{equation}
The classification of $(Q,S)$ sectors for decuplet-decuplet scattering and their corresponding flavor irreps involved can be found in Appendix~\ref{app1}.

Finally, some scattering sectors contain only one flavor irrep, corresponding to single-channel identical particle scatterings. For such scatterings, entanglement suppression exhibits a completely new and novel property, which we will discuss in Section~\ref{sec:subS}.

\section{Group-theoretic Identification of Symmetries}\label{sect:group}

Entanglement suppression has revealed relations between phase shifts, which implicitly hint at new symmetries. As demonstrated in Ref.~\cite{Liu:2022grf} for light octet baryons, these constraints manifest as emergent symmetries across different scattering sectors when implemented at the Lagrangian level. These symmetries can also be determined via group-theoretic counting by combining irreps of SU(2)$_{\rm spin}$ and SU(3)$_{\rm flavor}$ into irreps of the larger emergent symmetry, which provide further insights into the interplay between entanglement suppression and quantum statistics. We start by illustrating the counting in the spin-1/2 baryons and then apply it to the spin-3/2 cases.

\subsection{Spin-1/2 Octet Baryons}

Recall that octet baryons belong to the $\bm{8}$ irrep of the SU(3) flavor group. The spin and flavor structures of two octet baryons can be decomposed as
\begin{align}
&{\rm SU(2)}_{\rm spin}:\quad\,\frac12\otimes\frac12=\underbrace{0}_{\text{antisymmetric}}\oplus\underbrace{1}_{\text{symmetric}},\\
&{\rm SU(3)}_{\rm flavor}:\quad\bm{8}\otimes\bm{8}=\underbrace{\bm1\oplus\bm8_{\rm S}\oplus\bm{27}}_{\text{symmetric}}\oplus\underbrace{\bm8_{\rm A}\oplus\bm{10}\oplus\overline{\bm{10}}}_{\text{antisymmetric}}.
\end{align}
There is a single six-dimensional sector with $(Q,S)=(0,-2)$ in which scatterings involve all six flavor irreps. Minimizing the entanglement power in this sector leads to two enhanced symmetries~\cite{Liu:2022grf}: (i) an SU(16) spin-flavor symmetry associated with an Identity gate $S$-matrix, and (ii) an SU(8) flavor symmetry corresponding to a SWAP gate $S$-matrix. Let us consider them in turn.

When the $S$-matrix is an Identity gate one observes the universality of interaction strengths across all channels~\cite{Liu:2022grf},
\begin{equation}
      \delta_{\bm 1}=\delta_{{\bm 8}_{\rm S}}=\delta_{\bm{27}}=
     \delta_{{\bm 8}_{\rm A}}=\delta_{\bm{10}}=\delta_{\overline{\bm{10}}}\,,
\end{equation}
and a large SU(16) symmetry emerges, representing a full spin-flavor symmetry under which the eight baryons and the two spin states combine to form a $\bm{16}$, the fundamental representation. Hence the scattering of two spin-1/2 baryons in this case can be decomposed as
\begin{equation}
 {\rm SU(16)}_{\rm spin+flavor}:\quad\bm{16}\otimes \bm{16} = \underbrace{\bm{136}}_{\rm symmetric}\oplus \underbrace{\bm{120}}_{\rm antisymmetric}.
\end{equation}
FD statistics dictates that the scattering can only occur in the antisymmetric channel, the $\bm{120}$, which can be obtained by combining the symmetric $\bm{1}\oplus \bm{8}_{\rm S}\oplus \bm{27}$ with the $^1S_0$ spin singlet and the antisymmetric $\bm{8}_{\rm A}\oplus \bm{10}\oplus \overline{\bm{10}}$ with the $^3S_1$ spin triplet:
\begin{equation}
(1+8+27)\times1+(8+10+10)\times3=120\,.
\end{equation}
We emphasize that this is a consequence of the scattering dynamics being universal in all spin and flavor channels.

The SWAP gate is obtained when \cite{Liu:2022grf}
 \begin{equation}\label{eq:su8sol}
     \delta_{\bm 1}=\delta_{{\bm 8}_{\rm S}}=\delta_{\bm{27}}=\delta_{\rm S}\,, \qquad
     \delta_{{\bm 8}_{\rm A}}=\delta_{\bm{10}}=\delta_{\overline{\bm{10}}}=\delta_{\rm A}\,;\qquad |\delta_{\rm S} - \delta_{\rm A}|=\frac\pi2\,.
 \end{equation}
in which case all scattering phases within the symmetric or antisymmetric irreps are of equal strength but there is a difference between the symmetric and antisymmetric irreps. In this case Ref.~\cite{Liu:2022grf} identified an SU(8) flavor symmetry where all flavor-symmetric states combine into one single irrep and similarly for all flavor-antisymmetric states:
\begin{equation}
{\rm SU(8)}_{\rm flavor}:\quad\bm{8}\otimes \bm{8} = \underbrace{\bm{36}}_{\rm symmetric}\oplus \underbrace{\bm{28}}_{\rm antisymmetric},
\end{equation}
The counting works out as expected:
\begin{equation}
    1+8+27=36 \quad {\rm and} \quad 8+10+10=28 \,.
\end{equation}
The SU(8) flavor symmetry dictates that there are only two independent scattering channels: the flavor symmetric and the antisymmetric ones.

Next there are several three-dimensional sectors (e.g., $p\Lambda$-$p\Sigma^0$-$n\Sigma^+$) which scatter into  $\bm8_{\rm S}\oplus\bm{27}\oplus\bm8_{\rm A}\oplus\bm{10}\oplus\overline{\bm{10}}$ sectors. Entanglement suppression in the three-dimensional sectors arises from the conditions~\cite{Liu:2022grf}:
\begin{equation}
         \delta_{{\bm 8}_{\rm S}}=\delta_{\bm{27}}\equiv\delta_{\rm S}\,, \qquad
     \delta_{{\bm 8}_{A}}=\delta_{\bm{10}}=\delta_{\overline{\bm{10}}}\equiv\delta_{\rm A}\,;\qquad |\delta_{\rm S} - \delta_{\rm A}|=0\,, \quad {\rm or}\quad \frac\pi2\,.
\end{equation}
Since the $\bm1$ irrep is absent in this case, the numbers of symmetric and antisymmetric flavor states are
\begin{equation}
{\rm Symmetric}:~27+8 = 35 \,, \quad {\rm Antisymmetric}:~8+10+10 =28\,,
\end{equation}
It turns out that the SO(8) group has a rank-2 traceless symmetric irrep with dimension $35=8(8+1)/2-1$ and a rank-2 antisymmetric irrep with dimension $28=8(8-1)/2$, which match the counting above. Indeed, the SO(8) flavor symmetry was discovered in Ref.~\cite{Liu:2022grf} via the Lagrangian approach in the case of the Identity gate.  One can  verify that, in the case of the $S$-matrix being a SWAP gate, the Lagrangian also maintains the same SO(8) symmetry. 

\subsection{Spin-3/2 Decuplet Baryons}\label{sect:spin32}

Now we identify the possible emergent symmetries in the spin-3/2 baryon scatterings. In Section~\ref{sec:ES} we demonstrated that the $S$-matrix in the spin space is an Identity gate if all scattering phases are equal, as Eq.~\eqref{eq:deltasol1} shows. Again the universal strength of interactions in all channels suggests a large symmetry under which every scattering channel belongs to the same irrep. The total number of spin and flavor states is
\begin{equation}
    (27+28)\times (1+5) + (10+35)\times(3+7) = 780\,,
\end{equation}
where the number of spin states is given by $(2J+1)$, $J=0,1,2,3$. This number matches the dimension of the rank-2 antisymmetric representation of SU(40), arising from the product of two fundamental representations ($\bm{40}$):
\begin{equation}
 {\rm SU(40)}_{\rm spin+flavor}:\quad\bm{40}\otimes \bm{40}= \underbrace{\bm{820}}_{\rm symmetric} \oplus \underbrace{\bm{780}}_{\rm antisymmetric} .
\end{equation}
This indicates that the ten spin-3/2 baryons, each with four spin states, fully occupy the $\bm{40}$ of SU(40). Moreover, the $\bm{780}$ is precisely what the FD statistics demands, as the two-baryon system must be antisymmetric under state exchange!

On the other hand, the SWAP solution for the $S$-matrix in Eq.~\eqref{eq:deltasol2} requires the scattering dynamics to be universal only within the symmetric or antisymmetric flavor channels, which is similar to the spin-1/2 baryon case in Eq.~\eqref{eq:su8sol}. This suggests a symmetry group where all $55=27+28$ flavor symmetric states combine into one single irrep and similarly for all $45=10+35$ flavor antisymmetric states. Thus there is an enhanced SU(10) flavor symmetry:
\begin{equation}
 {\rm SU(10)}_{\rm flavor}:\quad \bm{10}\otimes \bm{10}= \underbrace{\bm{55}}_{\rm symmetric} \oplus \underbrace{\bm{45}}_{\rm antisymmetric}.
\end{equation}
As discussed in Section \ref{sec:theo}, the ability of the $S$-matrix to entangle spin states is correlated with its ability to entangle flavor states, due to quantum statistics. Thus, FD statistics dictates that an enhanced symmetry in flavor space implies an enhanced symmetry in spin space as well, which is manifest in the observation that $J=0$ and $J=2$ states now share the same scattering dynamics and similarly for $J=1$ and $J=3$ states. This results in an emergent SU(4)$_{\rm spin}$ symmetry, under which $J=0,2$ states form a six-dimensional rank-2 antisymmetric irrep and $J=1,3$ states form a ten-dimensional symmetric irrep:
\begin{equation}
\label{eq:su6106}
 {\rm SU(4)}_{\rm spin}:\quad \bm{4}\otimes \bm{4}= \underbrace{\bm{10}}_{\rm symmetric} \oplus \underbrace{\bm{6}}_{\rm antisymmetric} .
\end{equation}
In the end, the enhanced symmetry is $\text{SU}(4)_\text{spin}\times\text{SU}(10)_\text{flavor}$.

It is interesting to compare the SWAP solutions between spin-1/2 octet baryons and spin-3/2 decuplet baryons. In the former case the enhanced symmetry is SU(8)$_{\rm flavor}$ while in the latter case it is $\text{SU}(4)_\text{spin}\times\text{SU}(10)_\text{flavor}$. Again this is due to the fact that there is more than one single spin channel contributing to the symmetric or antisymmetric flavor states. As a result, the symmetry in the spin space is enhanced from SU(2) to SU(4). This emergent spin symmetry is not affected by whether exact flavor SU(3) symmetry is assumed, and it remains exact since its starting point, the spin SU(2) symmetry, is itself exact as long as no spin symmetry breaking effect, e.g., spin-orbital coupling, is considered. Considering a single scattering channel, e.g., $\Delta^0\Delta^+$, it will still exhibit an SU(4) spin symmetry after suppressing entanglement. The role of assuming exact flavor SU(3) symmetry is to relate the scattering in this channel to that in other channels, thereby allowing the emergence of a larger SU(40) spin-flavor symmetry. If flavor symmetry is not taken into account at all, then the emergent symmetry from entanglement suppression is simply the SU(4) spin symmetry (in this channel solely). Moreover, when scattering spin-$s$ baryons with $s>1/2$, we generally expect the spin symmetry to be enhanced to SU(2$s$+1)$_{\rm spin}$. We will see in Section~\ref{sec:subS} that the existence of two spin channels within the symmetric or antisymmetric flavor states also gives rise to a richer phenomenology when scattering two identical particles.

\section{Field-theoretic Identification of Symmetries}\label{sec:EFT}

As shown in Ref.~\cite{Liu:2022grf}, an explicit identification of emergent symmetries can be achieved in the Lagrangian approach. This requires establishing a connection between the phase shifts and the parameters in the Lagrangian. Since we are interested in the very low energy phenomenon, our description does not need to contain explicit pseudoscalar meson degrees of freedom. In this context, one often considers NREFT with short-range interactions, which can be matched to the effective range expansion (ERE)~\cite{Bethe:1949yr, Kaplan:1996xu, Hammer:2019poc}.

\subsection{NREFT for Decuplet Baryons}

Operators in an NREFT Lagrangian are ordered in powers of momenta or derivatives, and the LO Lagrangian contains only constant contact terms.
The LO Lagrangian for decuplet-decuplet interaction in NREFT reads as~\cite{Haidenbauer:2017sws}:
\begin{align}\label{eq:Lag4D}
	\mathcal{L} &=c_1\left( T_{abc}^{\dagger}T_{abc} \right) \left( T_{def}^{\dagger}T_{def} \right)\notag\\
	&\quad +c_2\left( T_{abc}^{\dagger}\Sigma ^{\alpha}T_{abc} \right) \left( T_{def}^{\dagger}\Sigma ^{\alpha}T_{def} \right)\notag\\
	&\quad +c_3\left( T_{abc}^{\dagger}\Sigma ^{\alpha \beta}T_{abc} \right) \left( T_{def}^{\dagger}\Sigma ^{\alpha \beta}T_{def} \right)\notag\\
	&\quad +c_4\left( T_{abc}^{\dagger}\Sigma ^{\alpha \beta \gamma}T_{abc} \right) \left( T_{def}^{\dagger}\Sigma ^{\alpha \beta \gamma}T_{def} \right)\notag\\
	&\quad +c_5\left( T_{abc}^{\dagger}T_{abd} \right) \left( T_{def}^{\dagger}T_{cef} \right)\notag\\
	&\quad +c_6\left( T_{abc}^{\dagger}\Sigma ^{\alpha}T_{abd} \right) \left( T_{def}^{\dagger}\Sigma ^{\alpha}T_{cef} \right)\notag\\
	&\quad +c_7\left( T_{abc}^{\dagger}\Sigma ^{\alpha \beta}T_{abd} \right) \left( T_{def}^{\dagger}\Sigma ^{\alpha \beta}T_{cef} \right)\notag\\
	&\quad +c_8\left( T_{abc}^{\dagger}\Sigma ^{\alpha \beta \gamma}T_{abd} \right) \left( T_{def}^{\dagger}\Sigma ^{\alpha \beta \gamma}T_{cef} \right),
\end{align}
where $c_i$'s are eight Wilson coefficients, also known as low-energy constants (LECs) in the context of low-energy effective field theories. Here, the decuplet baryons are represented by the totally symmetric three-index tensor $T$:
\begin{align}
	T_{111}&=\Delta ^{++},\quad T_{112}=\frac{1}{\sqrt{3}}\Delta ^+,\quad T_{122}=\frac{1}{\sqrt{3}}\Delta ^0,\quad T_{222}=\Delta ^-,\notag\\
	T_{113}&=\frac{1}{\sqrt{3}}\Sigma ^{*+},\quad T_{123}=\frac{1}{\sqrt{6}}\Sigma ^{*0},\quad T_{223}=\frac{1}{\sqrt{3}}\Sigma ^{*-},\notag\\
	T_{133}&=\frac{1}{\sqrt{3}}\Xi^{*0},\quad T_{233}=\frac{1}{\sqrt{3}}\Xi^{*-},\notag\\
	T_{333}&=\Omega ^-.
\end{align}
We also introduced unity, $\Sigma ^{\alpha}$, $\Sigma ^{\alpha\beta}$, and $\Sigma ^{\alpha\beta\gamma}$, corresponding to scalar, vector, rank-2, and rank-3 tensor operators, respectively. The operators $\Sigma ^{\alpha}/2$ are simply the generators of SU(2) in the four-dimensional representation: $\Sigma ^{\alpha}/2=t^\alpha_{s=3/2}$, and
\begin{align}
\Sigma ^{\alpha \beta}&=\frac{1}{8}\left( \Sigma ^{\alpha}\Sigma ^{\beta}+\Sigma ^{\beta}\Sigma ^{\alpha}-10\delta ^{\alpha \beta} \right) ,\notag\\
\Sigma ^{\alpha \beta \gamma}&=\frac{1}{36\sqrt{3}}\Big[ 5\big( \Sigma ^{\alpha}\Sigma ^{\beta}\Sigma ^{\gamma}+\Sigma ^{\gamma}\Sigma ^{\alpha}\Sigma ^{\beta}+\Sigma ^{\beta}\Sigma ^{\gamma}\Sigma ^{\alpha} +\Sigma ^{\alpha}\Sigma ^{\gamma}\Sigma ^{\beta}+\Sigma ^{\beta}\Sigma ^{\alpha}\Sigma ^{\gamma}+\Sigma ^{\gamma}\Sigma ^{\beta}\Sigma ^{\alpha} \big)\notag\\
&\quad -82\big( \Sigma ^{\alpha}\delta ^{\beta \gamma}+\Sigma ^{\gamma}\delta ^{\alpha \beta} +\Sigma ^{\beta}\delta ^{\gamma \alpha}\big)\Big].
\end{align}

It will be convenient to project the contact operators into irreps of SU(3) flavor symmetry, and the corresponding SU(3)-symmetric Wilson coefficients are~\cite{Haidenbauer:2017sws}
\begin{align}
&C_{0,\bm{27}}=\frac{1}{27}\big( -54c_1+810c_2-405c_3+6300c_4+6c_5-90c_6+45c_7-700c_8\big) ,\\
&C_{2,\bm{27}}=-2c_1+6c_2+9c_3+\frac{140}{3}c_4+\frac{2}{9}c_5-\frac{2}{3}c_6-c_7-\frac{140}{27}c_8\,,\\
&C_{0,\bm{28}}=\frac{1}{3}\big( -6c_1+90c_2-45c_3+700c_4-6c_5+90c_6-45c_7+700c_8 \big) ,\\
&C_{2,\bm{28}}=-2c_1+6c_2+9c_3+\frac{140}{3}c_4-2c_5+6c_6+9c_7+\frac{140}{3}c_8\,,\\
&C_{1,\overline{\bm{10}}}=\frac{1}{3}\big( -6c_1+66c_2-9c_3-420c_4+2c_5-22c_6+3c_7+140c_8 \big) ,\\
&C_{3,\overline{\bm{10}}}=-2c_1-18c_2-3c_3-\frac{20}{3}c_4+\frac{2}{3}c_5+6c_6+c_7+\frac{20}{9}c_8\,,\\
&C_{1,\bm{35}}=\frac{1}{3}\big( -6c_1+66c_2-9c_3-420c_4-2c_5+22c_6-3c_7-140c_8 \big) ,\\
&C_{3,\bm{35}}=-2c_1-18c_2-3c_3-\frac{20}{3}c_4-\frac{2}{3}c_5-6c_6-c_7-\frac{20}{9}c_8\,.
\end{align}

The ERE at LO~\cite{Bethe:1949yr, Kaplan:1996xu, Hammer:2019poc} establishes the relationship between the phase shift $\delta$ and the Wilson coefficient $C$ \cite{Liu:2022grf, Hu:2024hex}:
\begin{equation}\label{eq:ERE}
\lim_{p\to 0}\ p\cot \delta =-\frac1a=-\frac{2\pi}{\mu C}-\Lambda\,,
\end{equation}
where $p$ is the magnitude of the center-of-mass momentum, $a$ is the $S$-wave scattering length, $\mu=m/2$ is the reduced mass with $m$ being the decuplet baryon mass, and $\Lambda$ is the ultraviolet cutoff.

Utilizing Eq.~\eqref{eq:ERE}, we can deduce the Wilson coefficient $C$ for certain phase shift values. In particular, the non-interacting and unitarity limit cases are of great importance at LO:
\begin{align}
\label{eq:non-in}\delta=0:&\qquad a = 0\quad\,\ \text{or}\quad C=0\,,\\
\label{eq:uni}\delta=\frac\pi 2:&\qquad a=\infty\quad\text{or}\quad C=-\frac{2\pi}{\mu\Lambda}\,.
\end{align}
Both the free theory ($\delta=0$) and the unitarity limit ($\delta=\pi/2$) exhibit an enhanced symmetry: they are invariant under the Schrödinger group~\cite{Mehen:1999nd, Low:2021ufv}, the non-relativistic conformal group that leaves the Schrödinger equation invariant. Additionally, these two cases correspond to the fixed points of the renormalization group flow for non-relativistic two-body scattering governed by short-range interactions~\cite{Birse:1998dk}.

Furthermore, if we know the relation between the phase shifts of different channels, we can also use Eq.~\eqref{eq:ERE} to derive relations between the corresponding Wilson coefficients. In particular, entanglement suppression provides two relations between the phase shifts (denoting the two relevant channels as $\alpha$ and $\beta$):
\begin{equation}
\delta _\alpha=\delta _\beta:\quad a_\alpha=a_\beta\,,
\end{equation}
or
\begin{equation}
|\delta _\alpha-\delta _\beta|=\frac\pi2:\quad\frac{1}{a_\alpha a_\beta}=\lim_{p\to 0}\ p^2\cot \delta_\alpha\cot \delta_\beta=\lim_{p\to 0}\ (-p^2)\,.
\end{equation}
The former implies that the two channels have identical scattering lengths, meaning that their corresponding Wilson coefficients are equal at a given cutoff. The latter, in the limit of $p\to0$, indicates that one of the scattering lengths diverges (i.e., its phase shift reaches $\pi/2$), which in turn requires the phase shift of the other channel to vanish, and then their Wilson coefficients are given by Eqs.~\eqref{eq:non-in} and~\eqref{eq:uni}.

The symmetry of the Lagrangian can be made more transparent in one operator basis than in another. This will be helpful for analyzing emergent symmetries in the next subsections. Therefore, we rewrite the Lagrangian using an alternative set of operators:
\begin{align}\label{eq:Lag4Dnew}
	\mathcal{L} &=d_1\left( T_{abc}^{\dagger}T_{def}^{\dagger} \right) \left( T_{abc}T_{def} \right) \notag\\
	&\quad +d_2\left( T_{abc}^{\dagger}\Xi ^{\alpha *}T_{def}^{\dagger} \right) \left( T_{abc}\Xi ^{\alpha}T_{def} \right) \notag \\
	&\quad +d_3\left( T_{abc}^{\dagger}\Xi ^{\alpha \beta *}T_{def}^{\dagger} \right) \left( T_{abc}\Xi ^{\alpha \beta}T_{def} \right) \notag \\
	&\quad +d_4\left( T_{abc}^{\dagger}\Xi ^{\alpha \beta \gamma *}T_{def}^{\dagger} \right) \left( T_{abc}\Xi ^{\alpha \beta \gamma}T_{def} \right)\notag\\
	&\quad +d_5\left( T_{abc}^{\dagger}T_{def}^{\dagger} \right) \left( T_{abd}T_{cef} \right) \notag\\
	&\quad +d_6\left( T_{abc}^{\dagger}\Xi ^{\alpha *}T_{def}^{\dagger} \right) \left( T_{abd}\Xi ^{\alpha}T_{cef} \right) \notag \\
	&\quad +d_7\left( T_{abc}^{\dagger}\Xi ^{\alpha \beta *}T_{def}^{\dagger} \right) \left( T_{abd}\Xi ^{\alpha \beta}T_{cef} \right) \notag \\
	&\quad +d_8\left( T_{abc}^{\dagger}\Xi ^{\alpha \beta \gamma *}T_{def}^{\dagger} \right) \left( T_{abd}\Xi ^{\alpha \beta \gamma}T_{cef} \right),
\end{align}
where $d_i$'s are the LECs in the new operator basis, and the $\Xi$'s, along with their complex conjugates $\Xi^*$'s, are defined as
\begin{equation}
\Xi^\alpha=\exp \left( -i\pi \frac{\Sigma ^y}{2} \right)\Sigma^\alpha\,.
\end{equation}
In this new basis, the corresponding Wilson coefficients read as
\begin{align}
&C_{0,\bm{27}}=-8d_1+\frac{8}{9}d_5\,,&&C_{2,\bm{27}}=\frac{4}{3}(-9d_3+d_7),\\
&C_{0,\bm{28}}=-8(d_1+d_5),&&C_{2,\bm{28}}=-12(d_3+d_7),\\
&C_{1,\overline{\bm{10}}}=\frac{40}{3}(-3d_2+d_6),&&C_{3,\overline{\bm{10}}}=-\frac{400}{9}(3d_4-d_8),\\
&C_{1,\bm{35}}=-\frac{40}{3}(3d_2+d_6),&&C_{3,\bm{35}}=-\frac{400}{9}(3d_4+d_8).
\end{align}
In the following, we will switch between Eq.~\eqref{eq:Lag4D} and Eq.~\eqref{eq:Lag4Dnew} as needed.

\subsection{$\bm{27}\oplus\bm{28}\oplus\overline{\bm{10}}\oplus\bm{35}$ Sectors}

Having set up the theoretical framework, we can now analyze the emergent symmetries that entanglement suppression predicts in various sectors, starting from the global minimum in sectors containing all four flavor irreps $\bm{27}\oplus\bm{28}\oplus\overline{\bm{10}}\oplus\bm{35}$.

Let us first examine the sectors that involve all eight spin-flavor irreps. These include
\begin{align}
&\left(\Delta^-\Delta^{++}, \Delta^0\Delta^+\right),~\left(\Delta^-\Sigma^{*+}, \Delta^0\Sigma^{*0}, \Delta^+\Sigma^{*-}\right),\notag\\
&\left(\Delta^-\Xi^{*0}, \Delta^0\Xi^{*-}, \Sigma^{*-}\Sigma^{*0}\right),~\left(\Delta^+\Xi^{*-}, \Delta^0\Xi^{*0}, \Sigma^{*0}\Sigma^{*0}, \Sigma^{*+}\Sigma^{*-}\right),\notag\\
&\left(\Sigma^{*-}\Xi^{*-}, \Omega^-\Delta^-\right),~\left(\Sigma^{*-}\Xi^{*0}, \Sigma^{*0}\Xi^{*-}, \Omega^-\Delta^0\right), 
\end{align}
and their $I_3$-conjugate partners (see Appendix~\ref{app1}), when they exist.

The first non-entangling condition requires all eight phase shifts, and consequently all eight Wilson coefficients, to be equal, which sets $c_{2,\ldots,8}=0$. As a result, Eq.~\eqref{eq:Lag4D} reduces to
\begin{equation}
\mathcal{L} =c_1\left( T_{abc}^{\dagger}T_{abc} \right) \left( T_{def}^{\dagger}T_{def} \right),
\end{equation}
or equivalently, in the particle basis,
\begin{equation}
\mathcal{L} =c_1\left( \bm T^{\dagger i}\cdot\bm{T}^i \right)^2,
\end{equation}
where
\begin{equation}
\bm{T}=\left(\Delta ^{++},\Delta ^+,\Delta ^0,\Delta ^-,\Sigma ^{*+},\Sigma ^{*0},\Sigma ^{*-},\Xi^{*0},\Xi^{*-},\Omega ^-\right),
\end{equation}
and $i=1,2,3,4$ denotes the spin index. This Lagrangian exhibits a maximal SU(40) symmetry, which was identified using group-theoretic reasoning in Section \ref{sect:spin32}.

Another two minimal-entanglement solutions from Eq.~\eqref{eq:deltasol2} are
\begin{align}\label{eq:sol1}
&C_{0,\bm{27}}=C_{2,\bm{27}}=C_{0,\bm{28}}=C_{2,\bm{28}}=-\frac{2\pi}{\mu\Lambda}\,,\notag\\
&C_{1,\overline{\bm{10}}}=C_{3,\overline{\bm{10}}}=C_{1,\bm{35}}=C_{3,\bm{35}}=0\,,
\end{align}
and
\begin{align}\label{eq:sol2}
&C_{0,\bm{27}}=C_{2,\bm{27}}=C_{0,\bm{28}}=C_{2,\bm{28}}=0\,,\notag\\
&C_{1,\overline{\bm{10}}}=C_{3,\overline{\bm{10}}}=C_{1,\bm{35}}=C_{3,\bm{35}}=-\frac{2\pi}{\mu\Lambda}\,.
\end{align}
These two solutions correspond to the Lagrangians
\begin{align}\label{eq:Lagsol1}
	\mathcal{L} &=\frac{2\pi}{\mu\Lambda}\bigg[\frac{1}{4}\left( T_{abc}^{\dagger}T_{abc} \right) \left( T_{def}^{\dagger}T_{def} \right)-\frac{1}{4}\left( T_{abc}^{\dagger}T_{def} \right) \left( T_{def}^{\dagger}T_{abc} \right)\bigg]\notag\\
    &=\frac{2\pi}{\mu\Lambda}\bigg[\frac{1}{4}\left( \bm T^{\dagger i}\cdot\bm{T}^i \right)^2-\frac{1}{4}\left( \bm T^{\dagger i}\cdot\bm{T}^j \right)\left( \bm T^{\dagger j}\cdot\bm{T}^i \right)\bigg],
\end{align}
and
\begin{align}\label{eq:Lagsol2}
	\mathcal{L} &=\frac{2\pi}{\mu\Lambda}\bigg[\frac{1}{4}\left( T_{abc}^{\dagger}T_{abc} \right) \left( T_{def}^{\dagger}T_{def} \right)+\frac{1}{4}\left( T_{abc}^{\dagger}T_{def} \right) \left( T_{def}^{\dagger}T_{abc} \right)\bigg]\notag\\
    &=\frac{2\pi}{\mu\Lambda}\bigg[\frac{1}{4}\left( \bm T^{\dagger i}\cdot\bm{T}^i \right)^2+\frac{1}{4}\left( \bm T^{\dagger i}\cdot\bm{T}^j \right)\left( \bm T^{\dagger j}\cdot\bm{T}^i \right)\bigg],
\end{align}
respectively. A Fierz transformation has been employed to simplify these operators:
\begin{equation}
\frac14\delta_{ij}\delta_{kl}+\frac{1}{20}\left(\Sigma ^{\alpha}\right)_{ij}\left(\Sigma ^{\alpha}\right)_{kl}+\frac{1}{6}\left(\Sigma ^{\alpha \beta}\right)_{ij}\left(\Sigma ^{\alpha \beta}\right)_{kl}+\frac{3}{200}\left(\Sigma ^{\alpha \beta \gamma}\right)_{ij}\left(\Sigma ^{\alpha \beta \gamma}\right)_{kl}=\delta_{il}\delta_{kj}\,.
\end{equation}
Equations~\eqref{eq:Lagsol1} and~\eqref{eq:Lagsol2} exhibit an $\text{SU}(4)_\text{spin}\times\text{SU}(10)_\text{flavor}$ symmetry, in addition to the Schrödinger symmetry contained in Eqs.~\eqref{eq:sol1} and~\eqref{eq:sol2}. Moreover, observe that the $\left( \bm T^{\dagger i}\cdot\bm{T}^i \right)^2$ interaction preserves the spins of the incoming baryons after scattering, while  $\left( \bm T^{\dagger i}\cdot\bm{T}^j \right)\left( \bm T^{\dagger j}\cdot\bm{T}^i \right)$ interchanges the spins. In other words, Eqs.~\eqref{eq:Lagsol1} and~\eqref{eq:Lagsol2} are  proportional to 
\begin{equation}
\frac{1_\text{spin}-\operatorname{SWAP}_\text{spin}}{2}\,,\quad\text{and}\quad\frac{1_\text{spin}+\operatorname{SWAP}_\text{spin}}{2}\,,
\end{equation}
which are precisely the antisymmetric and symmetric projectors in spin space, respectively, again consistent with Eqs.~\eqref{eq:sol1} and~\eqref{eq:sol2}.

\subsection{Other Sectors}

In this subsection, we employ another set of LECs $d_i$, which will facilitate the identification of the emergent symmetry. The following sectors involve the $\bm{27}$, $\bm{28}$ and $\bm{35}$ irreps: $\left(\Delta^-\Delta^+, \Delta^0\Delta^0\right)$, $\left(\Delta^0\Sigma^{*-}, \Delta^-\Sigma^{*0}\right)$, $\left(\Delta^-\Xi^{*-}, \Sigma^{*-}\Sigma^{*-}\right)$, $\left(\Xi^{*-}\Xi^{*-}, \Omega^-\Sigma^{*-}\right)$, $\left(\Xi^{*-}\Xi^{*0}, \Omega^-\Sigma^{*0}\right)$ and their $I_3$-conjugate partners, if they exist. The phase-shift relations are given by
\begin{align}
&\delta_{0,\bm{27}}=\delta_{2,\bm{27}}=\delta_{0,\bm{28}}=\delta_{2,\bm{28}}\,,\qquad\delta_{1,\bm{35}}=\delta_{3,\bm{35}}\,;\\
&\delta_{0,\bm{27}}-\delta_{1,\bm{35}}=0\,,\quad\text{or}\quad\frac\pi2\,,\quad\text{or}\quad-\frac\pi2\,.\label{eq:3ch}
\end{align}
The Lagrangian corresponding to the case where $\delta_{0,\bm{27}}=\delta_{1,\bm{35}}$ is
\begin{align}
	\mathcal{L} &=d_1\left( T_{abc}^{\dagger}T_{def}^{\dagger} \right) \left( T_{abc}T_{def} \right)\notag\\
	&\quad +d_2\left( T_{abc}^{\dagger}\Xi ^{\alpha *}T_{def}^{\dagger} \right) \left( T_{abc}\Xi ^{\alpha}T_{def} \right)\notag\\
	&\quad +\frac23d_1\left( T_{abc}^{\dagger}\Xi ^{\alpha \beta *}T_{def}^{\dagger} \right) \left( T_{abc}\Xi ^{\alpha \beta}T_{def} \right)\notag\\
	&\quad +d_4\left( T_{abc}^{\dagger}\Xi ^{\alpha \beta \gamma *}T_{def}^{\dagger} \right) \left( T_{abc}\Xi ^{\alpha \beta \gamma}T_{def} \right)\notag\\
	&\quad +\left(\frac35d_1-3d_2\right)\left( T_{abc}^{\dagger}\Xi ^{\alpha *}T_{def}^{\dagger} \right) \left( T_{abd}\Xi ^{\alpha}T_{cef} \right)\notag\\
	&\quad +\left(\frac{9}{50}d_1-3d_4\right)\left( T_{abc}^{\dagger}\Xi ^{\alpha \beta \gamma *}T_{def}^{\dagger} \right) \left( T_{abd}\Xi ^{\alpha \beta \gamma}T_{cef} \right).
\end{align}
For $\delta_{0,\bm{27}}=\pi/2$ and $\delta_{1,\bm{35}}=0$, the Lagrangian becomes
\begin{align}
	\mathcal{L} &=\frac{\pi}{4\mu\Lambda}\left( T_{abc}^{\dagger}T_{def}^{\dagger} \right) \left( T_{abc}T_{def} \right)\notag\\
	&\quad +d_2\left( T_{abc}^{\dagger}\Xi ^{\alpha *}T_{def}^{\dagger} \right) \left( T_{abc}\Xi ^{\alpha}T_{def} \right)\notag\\
	&\quad +\frac{\pi}{6\mu\Lambda}\left( T_{abc}^{\dagger}\Xi ^{\alpha \beta *}T_{def}^{\dagger} \right) \left( T_{abc}\Xi ^{\alpha \beta}T_{def} \right)\notag\\
	&\quad +d_4\left( T_{abc}^{\dagger}\Xi ^{\alpha \beta \gamma *}T_{def}^{\dagger} \right) \left( T_{abc}\Xi ^{\alpha \beta \gamma}T_{def} \right)\notag\\
	&\quad -3d_2\left( T_{abc}^{\dagger}\Xi ^{\alpha *}T_{def}^{\dagger} \right) \left( T_{abd}\Xi ^{\alpha}T_{cef} \right)\notag\\
	&\quad -3d_4\left( T_{abc}^{\dagger}\Xi ^{\alpha \beta \gamma *}T_{def}^{\dagger} \right) \left( T_{abd}\Xi ^{\alpha \beta \gamma}T_{cef} \right).
\end{align}
Similarly, when $\delta_{0,\bm{27}}=0$ and $\delta_{1,\bm{35}}=\pi/2$, the Lagrangian takes the form
\begin{align}
	\mathcal{L} &=d_2\left( T_{abc}^{\dagger}\Xi ^{\alpha *}T_{def}^{\dagger} \right) \left( T_{abc}\Xi ^{\alpha}T_{def} \right)\notag\\
	&\quad +d_4\left( T_{abc}^{\dagger}\Xi ^{\alpha \beta \gamma *}T_{def}^{\dagger} \right) \left( T_{abc}\Xi ^{\alpha \beta \gamma}T_{def} \right)\notag\\
	&\quad +\left(\frac{3\pi}{20\mu\Lambda}-3d_2\right)\left( T_{abc}^{\dagger}\Xi ^{\alpha *}T_{def}^{\dagger} \right) \left( T_{abd}\Xi ^{\alpha}T_{cef} \right)\notag\\
	&\quad +\left(\frac{9\pi}{200\mu\Lambda}-3d_4\right)\left( T_{abc}^{\dagger}\Xi ^{\alpha \beta \gamma *}T_{def}^{\dagger} \right) \left( T_{abd}\Xi ^{\alpha \beta \gamma}T_{cef} \right).
\end{align}

In all three scenarios displayed above, it can be seen that at least two terms in the Lagrangian can be eliminated, indicating the presence of an enhanced symmetry. Only $d_1$, $d_2$, and $d_4$ remain as free parameters, and $d_1$ is further fixed if the interaction strength is specified in the latter two cases. An explicit identification of the symmetry in the Lagrangian remains to be further investigated. However, there are indications from the group-theoretic arguments presented in Section \ref{sect:group}. Notably, the flavor-symmetric (spin-antisymmetric) sector exhibits precisely the same symmetry structure as in the preceding subsection. This provides conclusive evidence for the existence of $\text{SU}(4)_\text{spin}\times\text{SU}(10)_\text{flavor}$ symmetry in the $\bm{27}\oplus\bm{28}$ channels. As for the $\bm{35}$ channels, an SU(4) spin symmetry can be verified, since $3+7=10$ spin states in this flavor irrep share the same coupling strength, thus forming a rank-2 symmetric representation of SU(4).

Finally, the following one-dimensional sectors involve the $\bm{28}$ and $\bm{35}$ irreps: $\Delta^-\Delta^0$, $\Delta^-\Sigma^{*-}$, $\Omega^-\Xi^{*-}$, and their $I_3$-conjugate partners. An SU(4) spin symmetry can always be concluded in these flavor irreps by counting numbers. Additional enlarged symmetries are suggested by relations between phase shifts or Wilson coefficients in the LO effective Lagrangian. However, identifying the underlying group-theoretic structure remains challenging.

\subsection{Large-$N_c$ Limit and Dibaryons}\label{sec:dis}

In the large-$N_c$ limit, where the number of colors $N_c\to\infty$, the intrinsic $\text{SU}(2)_\text{spin}\times\text{SU}(3)_\text{flavor}$ symmetry is enlarged to the SU(6) spin-flavor symmetry. In this scenario the spin-1/2 octet and spin-3/2 decuplet baryons combine to form a 56-dimensional representation described by a completely symmetric three-index field $\Psi_{\alpha\beta\gamma}$. The LO effective Lagrangian contains only two operators~\cite{Kaplan:1995yg}:
\begin{equation}\label{eq:su6Lag}
\mathcal{L}=-\left(\tilde{a}~\Psi^\dagger_{\alpha\beta\gamma}\Psi_{\alpha\beta\gamma}\Psi^\dagger_{\mu\nu\rho}\Psi_{\mu\nu\rho}+\tilde{b}~\Psi^\dagger_{\alpha\beta\gamma}\Psi_{\alpha\beta\rho}\Psi^\dagger_{\mu\nu\rho}\Psi_{\mu\nu\gamma}\right).
\end{equation}
In this limit of SU(6) spin-flavor symmetry, the eight Wilson coefficients are related to $\tilde{a}$ and $\tilde{b}$ in the above~\cite{Richardson:2024zln}:
\begin{align}\label{eq:su6Wil}
&C_{0,\bm{27}}
=C_{1,\overline{\bm{10}}}=2\left(\tilde{a}+\frac{\tilde{b}}{27}\right),\notag\\
&C_{2,\bm{27}}
=C_{1,\bm{35}}=2\left(\tilde{a}-\frac{\tilde{b}}{27}\right),\notag\\
&C_{0,\bm{28}}
=C_{3,\overline{\bm{10}}}=2\left(\tilde{a}-\frac{\tilde{b}}{3}\right),\notag\\
&C_{2,\bm{28}}
=C_{3,\bm{35}}=2\left(\tilde{a}+\frac{\tilde{b}}{3}\right).
\end{align}
We observe that each line in Eq.~\eqref{eq:su6Wil} equates one spin-symmetric component with one spin-antisymmetric component, in contrast to the global entanglement suppression prediction, for which
\begin{align}
&C_{0,\bm{27}}=C_{2,\bm{27}}=C_{0,\bm{28}}=C_{2,\bm{28}}\,,\notag\\
&C_{1,\overline{\bm{10}}}=C_{3,\overline{\bm{10}}}=C_{1,\bm{35}}=C_{3,\bm{35}}\,.
\label{eq:Cis}
\end{align}
The only solution that simultaneously satisfies both SU(6) symmetry and entanglement suppression is to choose $C_{0,\bm{27}}=C_{1,\overline{\bm{10}}}$, which immediately implies that all Wilson coefficients are equal. As can be directly seen from Eq.~\eqref{eq:su6Wil}, this corresponds to setting $\tilde{b}=0$. Once again, an emergent SU(40) symmetry appears in the decuplet-decuplet sector. 
In fact, the Lagrangian in Eq.~\eqref{eq:su6Lag} with $\tilde{b}=0$ also exhibits an SU(16) symmetry in the octet-octet sector, and an SU(56) symmetry when considering the full $\text{octet}+\text{decuplet}$ system---necessary for consistency in the large-$N_c$ limit. While an approximately vanishing $\tilde{b}$ has been observed in lattice calculations in the octet-octet sector~\cite{Wagman:2017tmp, NPLQCD:2020lxg}, such a result has not yet been seen in the sectors involving decuplet baryons.

Let us also discuss the possible $\Delta\Delta$ dibaryons in the framework of entanglement suppression. The $\Delta\Delta$ systems with $(J,I) = (3,0)$ and $(0,3)$ are in the irreps $(J,F)=(3,\overline{\bm{10}})$ and $(0,\bm{28})$, respectively, in the three-flavor case. The potential existence of the $(3,\overline{\bm{10}})$ resonance $d^*(2380)$ suggests that this channel approaches the unitarity limit, which might serve as an additional input for the sector $\left(\Delta^-\Delta^{++}, \Delta^0\Delta^+\right)$ that involves the flavor $\overline{\bm{10}}$ irrep. Demanding $\delta_{3,\overline{\bm{10}}}=\pi/2$ in this sector leads to two possibilities: (i) all eight irreps reach the unitarity limit, or (ii) spin-antisymmetric irreps are non-interacting and spin-symmetric irreps are at the unitarity limit, as shown in Eq.~\eqref{eq:sol2}. The former solution means that the $d^*(2380)$ is in the ${\bm{780}}$ multiplet under SU(40); in particular, this would predict the $d^*(2380)$ to have a partner of $(J,I) = (0,3)$.

\section{Identical Particles and Sub-unitary $S$-matrices}\label{sec:subS}

A special scenario which deserves a dedicated analysis is the scattering of two identical particles into a single flavor channel. For spin-1/2 octet baryons such a case includes $\{nn , pp, \Sigma^+\Sigma^+, \Sigma^-\Sigma^-,$ $\Xi^-\Xi^-,$ $\Xi^0\Xi^0\}$, all of which  scatter into the symmetric $\mathbf{27}$ irrep of SU(3)$_{\rm flavor}$. In this case FD statistics constrains the total angular momentum in the $S$-wave scattering to be in the antisymmetric spin singlet $^1S_0$ channel, which is already maximally entangled \cite{Liu:2022grf}.  The $S$-matrix contains a single channel:
\begin{equation}
\label{eq:nnS}
    \hat{S}_{\rm A} = {\cal J}_0\otimes {\cal F}_{\mathbf{27}}\,e^{2i\delta_{0,\mathbf{27}}},
\end{equation}
which is neither Identity nor SWAP. Notice that FD statistics restricts the physical states to the antisymmetric part of the total Hilbert space; in the symmetric part of the Hilbert space there can be neither free propagation nor scattering dynamics. Therefore, $\hat S_{\rm A}$ is not a unitary operator over the entire Hilbert space; instead it vanishes over the symmetric Hilbert (sub)space and is unitary only over the antisymmetric Hilbert (sub)space. In this case we call $\hat S_{\rm A}$ ``sub-unitary'' over the antisymmetric Hilbert space. Moreover,  the entanglement power of $\hat S_{\rm A}$ in Eq.~\eqref{eq:nnS} is constant, independent of the  scattering phase $\delta_{0,\mathbf{27}}$.

For spin-3/2 decuplet baryons, a more interesting scenario appears because both ${\cal J}=0$ and ${\cal J}=2$ are antisymmetric so there can be two different angular momentum channels, thereby allowing for interferences between the two when computing the entanglement power. For instance, the $(Q,S)=(-2,-6)$ sector, i.e., $\Omega\Omega$, scatters into a single flavor symmetric irrep $\bm{28}$ with two phase shifts $\delta_{0,\bm{28}}$ and $\delta_{2,\bm{28}}$:
\begin{equation}\label{eq:SOmegaOmega}
\hat S_{\rm A}=\mathcal{J}_{0}\,e^{2i\delta_{0,\bm{28}}}+\mathcal{J}_{2}\,e^{2i\delta_{2,\bm{28}}}.
\end{equation}
In this case the entanglement power is not a constant anymore and depends on the phase difference $\delta_{0,\bm{28}}-\delta_{2,\bm{28}}$. There are two other such channels for spin-2/3 baryons: $\Delta^-\Delta^-$ and $\Delta^{++}\Delta^{++}$, which are the $I_3$-conjugate of each other. A similar scenario arises in the spin-1 deuteron-deuteron ($dd$) scattering, which was studied in Ref.~\cite{Kirchner:2023dvg}. In what follows we will base our discussion on FD statistics; the generalization to BE statistics can be carried out trivially.

\subsection{Density Matrix and Entanglement Power for Identical Particles}

Formally speaking, the sub-unitary $S$-matrix can be viewed as a unitary $S$-matrix projected into the antisymmetric Hilbert space:
\begin{equation}
    \hat S_{\rm A} \equiv {\cal P}_{\rm A} \hat S\, {\cal P}_{\rm A} = \hat S\, {\cal P}_{\rm A} = {\cal P}_{\rm A} \, \hat S \,,\quad  \left[\hat S, {\cal P}_{\rm A}\right] = 0 \,,
\end{equation}
where ${\cal P}_{\rm A}$ is the projection operator in the antisymmetric Hilbert space defined in Eq.~\eqref{eq:SWAPdef}. Notice that $\hat S_{\rm A}^\dagger \hat S_{\rm A} = {\cal P}_{\rm A}$, implying unitarity over the antisymmetric Hilbert space, which is manifest when we go to a basis where ${\cal P}_{\rm A}$ is an Identity operator over the antisymmetric Hilbert space.\footnote{As mentioned in Section \ref{sec:theo}, such a basis consists of symmetric and antisymmetric base vectors.} It is easy to see that $\hat S_{\rm A}$ in Eq.~\eqref{eq:SOmegaOmega} indeed satisfies $\hat S_{\rm A}^\dagger \hat S_{\rm A}^{}=\mathcal{J}_{0}^{}+\mathcal{J}_{2}=\mathcal{P}_{\rm A}$. The resulting density matrix constructed out of $|\psi_{\rm out}\rangle_{\rm A} \equiv \hat S_{\rm A}|\psi_{\rm in}\rangle$ is not normalized to 1:
\begin{equation}
    \rho = |\psi_{\rm out}\rangle_{\rm A}\, _{\rm A}\langle \psi_{\rm out}| = \hat S_{\rm A} |\psi_{\rm in}\rangle\langle \psi_{\rm in}| \hat S_{\rm A}^\dagger \,, \quad {\rm Tr}[\rho]=_{\rm A}\!\!\langle \psi_{\rm out}|\psi_{\rm out}\rangle_{\rm A}= \langle \psi_{\rm in}|{\cal P}_{\rm A} |\psi_{\rm in}\rangle <  1\,.
\end{equation}
This situation is similar to the projection-valued measurement in quantum mechanics, in which case the properly normalized density matrix is \cite{Nielsen:2012yss}
\begin{equation}
\label{eq:modrho}
    \tilde{\rho}= \frac{{\cal P}_{\rm A}\, \rho \, {\cal P}_{\rm A}}{{\rm Tr}[\rho\, {\cal P}_{\rm A}]} = \frac{|\psi_{\rm out}\rangle_{\rm A}\, _{\rm A}\langle\psi_{\rm out}|}{_{\rm A}\langle\psi_{\rm out}|\psi_{\rm out}\rangle_{\rm A}}= \frac{\rho}{\langle \psi _{\mathrm{in}}|\mathcal P_{\rm A}|\psi _{\mathrm{in}}\rangle} \,.
\end{equation}
Then we have ${\rm Tr}\left[\tilde{\rho}\right]=1$.

The calculation of entanglement power for sub-unitary $S$-matrices also needs to be modified. The definition of entanglement power in Eq.~\eqref{eq:euave} employs an averaging procedure over the entire Hilbert space, which is clearly not applicable now given the restriction of the sub-unitary $S$-matrix to the antisymmetric Hilbert space. Instead, we propose that the entanglement power only averages over the physically accessible sub-Hilbert space, where the sub-unitary $S$-matrix projects into, and define the $k$-th-order-weighted entanglement power for linear entropy:
\begin{align}\label{eq:entid}
E_k(\hat S)&=\frac{1}{\int\mathrm{d}\omega_A\mathrm{d}\omega_B\langle \psi _{\mathrm{in}}|\mathcal P_{\rm A}|\psi _{\mathrm{in}}\rangle^k} \int\mathrm{d}\omega_A\mathrm{d}\omega_B\langle \psi _{\mathrm{in}}|\mathcal P_{\rm A}|\psi _{\mathrm{in}}\rangle^k\Big(1-\operatorname{Tr}_A\left[\tilde{\rho}_A^2\right]\Big) \notag\\ 
&=1-\frac{\int\mathrm{d}\omega_A\mathrm{d}\omega_B\langle \psi _{\mathrm{in}}|\mathcal P_{\rm A}|\psi _{\mathrm{in}}\rangle^k\,\operatorname{Tr}_A\left[\tilde{\rho}_A^2\right]}{\int\mathrm{d}\omega_A\mathrm{d}\omega_B\langle \psi _{\mathrm{in}}|\mathcal P_{\rm A}|\psi _{\mathrm{in}}\rangle^k}\,.
\end{align}
It can be verified that the above expression satisfies all essential conditions required for an ``entanglement power''. For example, when the two subsystems have the same dimension $n$, the original entanglement power ranges from 0 to $1-1/n$~\cite{Chang:2024wrx}, and the values of $E_k$ also fall within this range. Additionally, for neutron-neutron ($nn$) scattering, $E_k(\hat S)=1/2$ for any $k$ (see Appendix~\ref{app2} for a proof), which is precisely what we expect given that the $S$-wave $nn$ system is always in a maximally entangled $^1S_0$ spin singlet state.

It is computationally straightforward to calculate $E_k$ with $k=2$, as the $\langle \psi _{\mathrm{in}}|\mathcal P_{\rm A}|\psi _{\mathrm{in}}\rangle^2$ factors cancel out the denominator in Eq.~\eqref{eq:modrho}:
\begin{equation}
\label{eq:entpowid}
E_2(\hat S)=1-\frac{\int\mathrm{d}\omega_A\mathrm{d}\omega_B\operatorname{Tr}_A\left[\rho_A^2\right]}{\int\mathrm{d}\omega_A\mathrm{d}\omega_B\langle \psi _{\mathrm{in}}|\mathcal P_{\rm A}|\psi _{\mathrm{in}}\rangle^2}\,,
\end{equation}
using which we can compute the entanglement power  for the $\Omega\Omega$ $S$-matrix in Eq.~\eqref{eq:SOmegaOmega}:
\begin{equation}
\label{eq:entnon}
E_2(\hat S)=\frac{1}{48}\big\{25-\cos\left[4\left(\delta_{0,\bm{28}}-\delta_{2,\bm{28}}\right)\right]\big\}.
\end{equation}
We find that $E_2(\hat S)$ never vanishes, since the $\Omega\Omega$ state itself is highly entangled already. Nevertheless, there are two minima, which take place  when
\begin{equation}\label{eq:oodelta}
\left|\delta_{0,\bm{28}}-\delta_{2,\bm{28}}\right|=0\,,\quad\text{or}\quad\frac\pi2\,.
\end{equation}
Since the lattice result for the scattering length in $^1S_0$ channel of $\Omega\Omega$ system is rather large, about 4.6 fm at a nearly physical pion mass $m_\pi\simeq 146$~MeV~\cite{Gongyo:2017fjb}, it may suggest $\delta_{0,\bm{28}}=\pi/2$, so $\delta_{2,\bm{28}}=0$ or $\pi/2$.

\subsection{Identity and SWAP Gates for Identical Particles}

Let us interpret the two minima found in Eq.~\eqref{eq:oodelta}. Substituting $\delta_{0,\bm{28}} = \delta_{2,\bm{28}}$ into Eq.~\eqref{eq:SOmegaOmega} yields
\begin{equation}
   \hat S_{\rm A} \propto \mathcal{J}_{2} + \mathcal{J}_{0} = {\cal P}_{\rm A}\,, 
\end{equation}
which is the projection operator onto antisymmetric states $\sum_{i}\mathcal{A}_{i}$ and, as commented previously, can be interpreted as the Identity operator $1_{\rm A}$ in the sub-Hilbert space spanned by antisymmetric states. On the other hand, substituting $\delta_{0,\bm{28}} = \pi/2$ and $\delta_{2,\bm{28}} = 0$, we obtain
\begin{equation}
   \hat S_{\rm A} = \mathcal{J}_{2} - \mathcal{J}_{0} \equiv \operatorname{SWAP}_{\rm A} \,,
\end{equation}
which is the analog of the SWAP gate acting on the antisymmetric Hilbert space. Observe that $\left(\operatorname{SWAP}_{\rm A}\right)^2={\cal P}_{\rm A}=1_{\rm A}$, just like $(\operatorname{SWAP})^2=1$ for the unrestricted Hilbert space.

These considerations lead to
\begin{equation}
   \mathcal{J}_{0} = \frac{{\cal P}_{\rm A} - \operatorname{SWAP}_{\rm A}}{2}\,, \quad
   \mathcal{J}_{2} = \frac{{\cal P}_{\rm A} + \operatorname{SWAP}_{\rm A}}{2}\,.
\end{equation}
Moreover, $|J=0\rangle$ and $|J=2\rangle$ states are eigenstates of $\operatorname{SWAP}_{\rm A}$ with $\mp 1$ eigenvalues:
\begin{equation}
    \operatorname{SWAP}_{\rm A}|J=2\rangle = |J=2\rangle \,, \quad  \operatorname{SWAP}_{\rm A}|J=0\rangle = -|J=0\rangle \,,
\end{equation}
which is parallel to the definition of SWAP in Eq.~\eqref{eq:swapdef} for the unrestricted Hilbert space. The analogy can be developed further by considering the $J_z=0$ states in both $J=0$ and $J=2$:
\begin{align}
   \ket{\alpha} &= \frac{1}{\sqrt{2}}\left(\ket{J=2;J_z=0} + \ket{J=0;J_z=0}\right), \\
   \ket{\beta} &= \frac{1}{\sqrt{2}}\left(\ket{J=2;J_z=0} - \ket{J=0;J_z=0}\right),
\end{align}
then the action of $\operatorname{SWAP}_{\rm A}$ becomes
\begin{equation}
   \operatorname{SWAP}_{\rm A} \ket{\alpha} = \ket{\beta}\,, \quad
   \operatorname{SWAP}_{\rm A} \ket{\beta}  = \ket{\alpha}\,,
\end{equation}
which indicates that it exchanges the states $\ket{\alpha}$ and $\ket{\beta}$.

Notice that it is the existence of two scattering channels in $J=0$ and $J=2$ which makes it possible to define an operator like $\operatorname{SWAP}_{\rm A}$ in two-qudit scattering. For a two-qubit system such as the $nn$ scattering, only one channel exists and one cannot introduce the $\operatorname{SWAP}_{\rm A}$ operator.

\subsection{Emergent Symmetries for Identical Particles}

It is important to point out that the solution in Eq.~\eqref{eq:oodelta} indeed yields enhanced symmetries. When the $S$-matrix is ${\mathcal P}_{\rm A}$, which is the Identity gate over the antisymmetric Hilbert space, there is a ``spin symmetry'' since $\delta_{0,\bm{28}}=\delta_{2,\bm{28}}$ and the scattering dynamics is independent of the total spin of the system.

The symmetry group can be identified as follows. As the spin of the $\Omega$ baryon is $3/2$, there are four spin states with $J_z=\pm 1/2,\pm 3/2$ and the spin symmetry group is SU(4) under which the four states form the fundamental representation $\bm{4}$. Then the relation $\delta_{0,\bm{28}}=\delta_{2,\bm{28}}$ is enforced by the SU(4) spin symmetry as the $J=0$ and $J=2$ states now furnish the rank-2 antisymmetric irrep $\bm{6}$. Out of the two possibilities shown in Eq.~\eqref{eq:su6106}, $\bm{6}$ is realized because of FD statistics. On the other hand, Schrödinger symmetry emerges when the $S$-matrix is ${\rm SWAP}_{\rm A}$, $|\delta_{0,\bm{28}}-\delta_{2,\bm{28}}|=\pi/2$ and $\delta_{0,\bm{28}}=\pi/2$ as suggested by the lattice result in Ref.~\cite{Gongyo:2017fjb}. In other words, although the entanglement power in Eq.~\eqref{eq:entnon} never vanishes for identical particle scattering, there are still enhanced symmetries when the entanglement power is minimized.

A similar conclusion can be reached in the case of spin-1 $dd$ scattering, which was first considered in Ref.~\cite{Kirchner:2023dvg}. Due to BE statistics, the $S$-wave $dd$ scattering has total angular momentum $J=0$ and $J=2$, which is similar to the two spin channels in identical spin-3/2 $\Omega\Omega$ scattering (but here $J=0$ and $J=2$ are symmetric components). Using Eq.~\eqref{eq:entpowid}, the entanglement power for $dd$ scattering can be written in terms of the phase difference in the $J=0$ and $J=2$ channel:\footnote{The denominator in Eq.~\eqref{eq:entpowid} is different from that used in Ref.~\cite{Kirchner:2023dvg} to compute the entanglement power in the spin-1 $dd$ scattering, where the corresponding denominator is $\left(\int\mathrm{d}\omega_A\mathrm{d}\omega_B\langle \psi _{\mathrm{in}}|\mathcal P_{\rm S}|\psi _{\mathrm{in}}\rangle\right)^2$. We thank Hans-Werner Hammer for clarifications on this point.}
\begin{equation}
E_2(\hat S)=\frac{171-70\cos \left[ 2\left( \delta _0-\delta _2 \right) \right] -20\cos \left[ 4\left( \delta _0-\delta _2 \right) \right]}{594}\,.
\end{equation}
Notice the dependence on both $2(\delta_0-\delta_2)$ and $4(\delta_0-\delta_2)$. Consequently the entanglement power has a global minimum at $\delta_0=\delta_2$ and a local minimum at $|\delta_0-\delta_2|=\pi/2$. This is in contrast with the entanglement power for spin-3/2 $\Omega\Omega$ scattering in Eq.~\eqref{eq:entnon}, which only depends on $4(\delta_{0,\bm{28}}-\delta_{2,\bm{28}})$ and both solutions, $|\delta_{0,\bm{28}}-\delta_{2,\bm{28}}|=0$ or $\pi/2$, are global minima. It is intriguing that ${\rm SWAP}_{\rm S}$ corresponds to the local minimum in $dd$ scattering while ${\cal P}_{\rm S}$ gives the global minimum. How general this observation is remains to be investigated further.

Following the discussion of enhanced symmetries in $\Omega\Omega$ scattering, the enhanced spin symmetry in $dd$ scattering for $\delta_0=\delta_2$ is SU(3)$_{\rm spin}$, under which the $J_z=0,\pm 1$ states form a $\bm{3}$ of SU(3) and the group theory now dictates:
\begin{equation}
    {\rm SU(3)_{\rm spin}}:\quad\bm{3}\otimes \bm{3} = \underbrace{{\bm 6}}_{\rm symmetric}\oplus \underbrace{{\bm 3}}_{\rm antisymmetric}.
\end{equation}
The symmetric $\bm{6}$ is realized in $dd$ scattering because of BE statistics. The other scenario $|\delta_0-\delta_2|=\pi/2$ can be realized if one of the channels flows to the unitarity limit and exhibits  Schrödinger symmetry, as discussed previously.

More generally, when scattering two identical spin-$s$ particles in the $S$-wave, quantum statistics determines that the total angular momentum $J$ can only take even values for both fermions and  bosons.\footnote{Recall that when adding two identical half-integral angular momenta, the even total angular momentum is antisymmetric under the exchange of two particles. On the other hand, when adding two identical integral angular momenta, the symmetric states also have even total angular momentum!} According to the general result in Section \ref{sec:theo}, the $S$-matrix is an Identity gate when all scattering phases are equal. In this case, group-theoretic counting indicates that there is an enhanced SU(2$s$+1)$_{\rm spin}$ symmetry, under which the (2$s$+1) spin states furnish a fundamental representation. In this case,
\begin{equation}
    {\rm{SU}}(2s+1)_{\rm spin}:\quad(\bm{2s+1})\otimes (\bm{2s+1}) = \underbrace{(\bm{s+1})(\bm{2s+1})}_{\rm symmetric}\oplus \underbrace{\bm s(\bm{2s+1})}_{\rm antisymmetric}.
\end{equation}
The dimensions of rank-2 symmetric and antisymmetric irreps agree with the numbers of symmetric and antisymmetric $J$ states, respectively, 
\begin{equation}
\sum_{\substack{{\rm even} \ J,\\ J\le 2s}}(2J+1) = \begin{cases} 
(s+1)(2s+1)\,, \quad \mbox{for integral}\ s\,,\\
s(2s+1) \,, \qquad \quad\ \,\mbox{for half-integral}\ s\,. 
\end{cases}
\end{equation}

\section{Summary and Outlook}\label{sec:sum}

In this paper, we explored the consequence of the conjectured entanglement suppression in the 2-to-2 $S$-wave scattering of spin-3/2 baryons, which are four-dimensional qudits in the spin space. This hypothesis posits that under certain conditions, low-energy scatterings minimize entanglement in the spin degrees of freedom, which can lead to the emergence of enhanced symmetries. Spin-3/2 baryons transform as a ten-dimensional irrep under the light-flavor SU(3) symmetry and are considered indistinguishable under strong interactions. Our analysis places particular emphasis on the interplay between spin and flavor symmetries due to the constraints of quantum statistics.

Using the general result that, for a bipartite system of equal dimensions, entanglement suppression takes place only for the Identity and SWAP gates, we presented the conditions on the scattering phases leading to entanglement suppression in spin-3/2 decuplet scattering. Using group-theoretic arguments, we identify a large SU(40)$_{\rm spin+flavor}$ when the $S$-matrix is an Identity gate, as well as an $\text{SU}(4)_\text{spin}\times\text{SU}(10)_\text{flavor}$ symmetry combined with the Schrödinger symmetry in the case of the SWAP gate. These observations are subsequently verified using non-relativistic effective field theories for spin-3/2 baryons. These findings generalize the results from the scattering of spin-1/2 octet baryons.

We also carefully examine the scattering of identical particles and the related computation of entanglement power. While this case has been considered previously in the literature, there are subtleties arising from the fact that quantum statistics restricts that only the symmetric or antisymmetric portion of the two-particle Hilbert space is physically accessible, which needs to be dealt with deliberately. In particular, we propose that the entanglement power only averages over the symmetric or antisymmetric Hilbert space, depending on whether the interaction satisfies the BE or FD statistics, and proceed to demonstrate that one can define the corresponding Identity and SWAP gates for identical particle scattering. Furthermore, even though the entanglement power is generally non-vanishing due to quantum statistics, enhanced symmetries still appear at the global or local minimum. For the scattering of identical spin-$s$ particles, we identify the enhanced symmetry groups to be SU$(2s+1)_{\rm spin}$ and/or the non-relativistic conformal group.

It is worth emphasizing that the scattering dynamics of systems beyond two identical qubits contains a rich phenomenology and offers important insights into the interplay between entanglement suppression, enhanced symmetries and quantum statistics. This is because in two-qubit systems the $S$-matrix in the symmetric or antisymmetric Hilbert space only contains one single scattering channel in the spin space, which renders the entanglement power constant. It is the existence of two or more scattering channels which results in the possibility of global or local minimum in the non-vanishing entanglement power. In the end, the entanglement power does not have to vanish to lead to enhanced symmetries. A global or local minimum is sufficient.

As a natural continuation of the current framework, octet-decuplet scattering  involves a bipartite system with differing dimensions for the subsystems. This scenario will be studied in the future.

\begin{acknowledgments}

We would like to thank Hao-Jie Jing for discussions regarding the generators of SU(3) irreps and the construction of Lagrangian.
This work was initiated during the long-term workshop, HHIQCD2024, at the Yukawa Institute for Theoretical Physics (YITP-T-24-02) in Kyoto University and we thank the workshop organizers and the participants for a stimulating meeting.
TRH is supported in part by the China National Training Program of Innovation and Entrepreneurship for Undergraduates under Grant No.~202414430004;
KS and TH are supported by the Grants-in-Aid for Scientific Research from JSPSs (Grants No.~JP23H05439, and No.~JP22K03637); and by JST, the establishment of university fellowships towards the creation of science technology innovation, Grant No.~JPMJFS2123;
TRH and FKG are supported in part by the National Natural Science Foundation of China under Grants No. 12125507, No. 12361141819, and No. 12447101; and by the Chinese Academy of Sciences under Grants No.~YSBR-101;
IL is supported in part by the U.S. Department of Energy under contracts DE-AC02-06CH11357, DE-SC0023522, DE-SC0010143  and No. 89243024CSC000002 (QuantISED Program).

\end{acknowledgments}

\appendix

\section{Classification of Decuplet-Decuplet Channels}\label{app1}

The channels of the decuplet-decuplet scattering and their charge $Q$, total isospin $I$, total spin $J$ and corresponding flavor irreps $F$ in the $S$-wave scattering are summarized in Table~\ref{tbl:S=0 secters}-\ref{tbl:S=-6 secters}. Each table is classified in terms of the strangeness (hypercharge) $S$ ($Y$) of the system, and therefore the $(Q,S)$ sectors can be found from the tables. The spin-antisymmetric (spin-symmetric) component $J_{\rm A}=0,2$ ($J_{\rm S}=1,3$) and their corresponding flavor-symmetric (flavor-antisymmetric) irreps $F_{\rm S}=\bm{27},\bm{28}$ ($F_{\rm A}=\overline{\bm{10}},\bm{35}$) are shown in the $[J_{\rm A},F_{\rm S}]$ ($[J_{\rm S},F_{\rm A}]$) column in each table. The isospin, spin, and flavor irreps in the lower half of each table are omitted, as they are identical to those of the $I_3$-conjugate channels presented in the upper half.

The flavor irrep $\overline{\bm{10}}$ does not contain the $S=-4,-5$ and $-6$ sectors, and therefore the channels in Tables~\ref{tbl:S=-4 secters},\ref{tbl:S=-5 secters} and \ref{tbl:S=-6 secters} are within the  $\bm{27},\bm{28}$ and $\bm{35}$ irreps. Moreover, in the case of $S=-6$, the channel with $\Omega\Omega$ belongs to only $\bm{28}$ and their spin components are restricted to antisymmetric ones $J=0,2$ as discussed in Section~\ref{sec:subS}. On the other hand, the channels including all possible flavor irreps  can be found in the $S=0,-1-2$ and $-3$ sectors.

\begin{table}[tbp] 
\caption{The decuplet-decuplet scattering channels and their charge $Q$, total isospin $I$, total spin $J$ and corresponding flavor irreps $F$ for the strangeness (hypercharge) $S=0$ ($Y=2$) sector. Channels in the bottom half  are the $I_3$-conjugate of those in the upper half and have the same spin and flavor irreps. \label{tbl:S=0 secters}}
\begin{ruledtabular}
\begin{tabular}{lcccc}
    Channels for $S=0$ & $Q$ & $I$ 
    & $[J_{\rm A},F_{\rm S}]$ & $[J_{\rm S},F_{\rm A}]$   \\
    \hline
    $\Delta^-\Delta^-$ & $-2$ & $3$ 
    & $[0,\bm{28}],[2,\bm{28}]$ &\\
    $\Delta^-\Delta^0$ & $-1$ & $2,3$ 
    & $[0,\bm{28}],[2,\bm{28}]$ & $[1,\bm{35}],[3,\bm{35}]$ \\
     $\Delta^-\Delta^+$, $\Delta^0\Delta^0$
     & $0$ & $1,2,3$ & $[0,\bm{27}],[0,\bm{27}],[2,\bm{28}],[2,\bm{28}]$ 
     & $[1,\bm{35}],[3,\bm{35}]$   \\
         $\Delta^-\Delta^{++}$, $\Delta^0\Delta^+$
     & $1$ & $0,1,2,3$ & $[0,\bm{27}],[0,\bm{27}],[2,\bm{28}],[2,\bm{28}]$ 
     & $[1,\overline{\bm{10}}],[1,\overline{\bm{10}}],[1,\bm{35}],[3,\bm{35}]$  \\
     $\Delta^0\Delta^{++}$, $\Delta^+\Delta^+$ & $2$ & \\
    $\Delta^+\Delta^{++}$ & $3$ &  \\
    $\Delta^{++}\Delta^{++}$ & $4$ & \\
\end{tabular}
\end{ruledtabular}
\end{table}

\begin{table}[tbp]
\caption{Same with Table~\ref{tbl:S=0 secters} but for the $S=-1$ ($Y=1$) sector. \label{tbl:S=-1 secters}}
\begin{ruledtabular}
    \begin{tabular}{lcccc}
    Channels for $S=-1$ & $Q$ & $I$
    & $[J_{\rm A},F_{\rm S}]$ & $[J_{\rm S},F_{\rm A}]$\\
    \hline
    $\Delta^-\Sigma^{*-}$ & $-2$ & $5/2$ & $[0,\bm{28}],[2,\bm{28}]$ 
    & $[1,\bm{35}],[3,\bm{35}]$  \\
    $\Delta^0\Sigma^{*-}$, $\Delta^-\Sigma^{*0}$
     & $-1$ & $3/2,5/2$ & $[0,\bm{27}],[2,\bm{27}],[0,\bm{28}],[2,\bm{28}]$ 
     & $[1,\bm{35}],[3,\bm{35}]$ \\
     $\Delta^-\Sigma^{*+}$,$\Delta^0\Sigma^{*0}$,$\Delta^+\Sigma^{*-}$ 
     & $0$ & $1/2,3/2,5/2$ & $[0,\bm{27}],[2,\bm{27}],[0,\bm{28}],[2,\bm{28}]$ 
     & $[1,\overline{\bm{10}}],[3,\overline{\bm{10}}],[1,\bm{35}],[3,\bm{35}]$  \\ 
     $\Delta^0\Sigma^{*+}$,$\Delta^+\Sigma^{*0}$,$\Delta^{++}\Sigma^{*-}$
     & $1$ & &   \\
     $\Delta^+\Sigma^{*+}$,$\Delta^{++}\Sigma^{*0}$
     & $2$ &  &   \\
    $\Delta^{++}\Sigma^{*+}$ & $3$ &  &   
\end{tabular}
\end{ruledtabular}
\end{table}

\begin{table}[tbp]
\caption{Same with Table~\ref{tbl:S=0 secters} but for the $S=-2$ ($Y=0$) sector. \label{tbl:S=-2 secters}}
\begin{ruledtabular}
    \begin{tabular}{lcccc}
    Channels for $S=-2$ & $Q$ & $I$ 
    & $[J_{\rm A},F_{\rm S}]$ & $[J_{\rm S},F_{\rm A}]$  \\
    \hline
    $\Delta^-\Xi^{*-}$, $\Sigma^{*-}\Sigma^{*-}$ & $-2$ & $2$ 
    & $[0,\bm{27}],[2,\bm{27}],[0,\bm{28}],[2,\bm{28}]$ 
     & $[1,\bm{35}],[3,\bm{35}]$\\
     $\Delta^-\Xi^{*0}$, $\Delta^0\Xi^{*-}$, $\Sigma^{*-}\Sigma^{*0}$
     & $-1$ & $1,2$ & $[0,\bm{27}],[2,\bm{27}],[0,\bm{28}],[2,\bm{28}]$ 
     & $[1,\overline{\bm{10}}],[3,\overline{\bm{10}}],[1,\bm{35}],[3,\bm{35}]$\\
      $\Delta^+\Xi^{*-}$, $\Delta^0\Xi^{*0}$, $\Sigma^{*0}\Sigma^{*0}$, 
      $\Sigma^{*+}\Sigma^{*-}$
     & $0$ & $0,1,2$ &  $[0,\bm{27}],[2,\bm{27}],[0,\bm{28}],[2,\bm{28}]$ 
     & $[1,\overline{\bm{10}}],[3,\overline{\bm{10}}],[1,\bm{35}],[3,\bm{35}]$  \\
         $\Delta^+\Xi^{*0}$,$\Delta^{++}\Xi^{*-}$, $\Sigma^{*+}\Sigma^{*0}$
     & $1$  &  \\
     $\Delta^{++}\Xi^{*0}$, $\Sigma^{*+}\Sigma^{*+}$
     & $2$ & \\
\end{tabular}
\end{ruledtabular}
\end{table}

\begin{table}[tbp]
\caption{Same with Table~\ref{tbl:S=0 secters} but for the $S=-3$ ($Y=-1$) sector. \label{tbl:S=-3 secters}}
\begin{ruledtabular}
    \begin{tabular}{lcccc}
    Channels for $S=-3$ & $Q$ & $I$ 
    & $[J_{\rm A},F_{\rm S}]$ & $[J_{\rm S},F_{\rm A}]$   \\
    \hline
    $\Sigma^{*-}\Xi^{*-}$, $\Omega^- \Delta^-$
    & $-2$ & $3/2$ &  $[0,\bm{27}],[2,\bm{27}],[0,\bm{28}],[2,\bm{28}]$ 
     & $[1,\overline{\bm{10}}],[3,\overline{\bm{10}}],[1,\bm{35}],[3,\bm{35}]$\\
    $\Sigma^{*-}\Xi^{*0}$, $\Sigma^{*0}\Xi^{*-}$, $\Omega^- \Delta^0$
    & $-1$ & $1/2,3/2$ & $[0,\bm{27}],[2,\bm{27}],[0,\bm{28}],[2,\bm{28}]$ 
     & $[1,\overline{\bm{10}}],[3,\overline{\bm{10}}],[1,\bm{35}],[3,\bm{35}]$    \\
    $\Sigma^{*0}\Xi^{*0}$, $\Sigma^{*+}\Xi^{*-}$, $\Omega^- \Delta^+$ 
    & $0$ &    \\
    $\Sigma^{*+}\Xi^{*0}$, $\Omega^- \Delta^{++}$
    & $1$ &  \\
\end{tabular}
\end{ruledtabular}
\end{table}

\begin{table}[tbp]
\caption{Same with Table~\ref{tbl:S=0 secters} but for the $S=-4$ ($Y=-2$) sector. \label{tbl:S=-4 secters}}
\begin{ruledtabular}
    \begin{tabular}{lcccc}
    Channels for $S=-4$ & $Q$ & $I$ & $[J_{\rm A},F_{\rm S}]$ & $[J_{\rm S},F_{\rm A}]$   \\
    \hline
    $\Xi^{*-}\Xi^{*-}$, $\Omega^-\Sigma^{*-}$
    & $-2$ & $1$ & $[0,\bm{27}],[2,\bm{27}],[0,\bm{28}],[2,\bm{28}]$ 
     & $[1,\bm{35}],[3,\bm{35}]$  \\
    $\Xi^{*-}\Xi^{*0}$, $\Omega^-\Sigma^{*0}$
    & $-1$ & $0,1$ & $[0,\bm{27}],[2,\bm{27}],[0,\bm{28}],[2,\bm{28}]$ 
    &  $[1,\bm{35}],[3,\bm{35}]$  \\
    $\Xi^{*0}\Xi^{*0}$, $\Omega^-\Sigma^{*+}$
    & $0$  &   \\
\end{tabular}
\end{ruledtabular}
\end{table}

\begin{table}[tbp]
\caption{Same with Table~\ref{tbl:S=0 secters} but for the $S=-5$ ($Y=-3$) sector. \label{tbl:S=-5 secters}}
\begin{ruledtabular}
    \begin{tabular}{lcccc}
    Channels for $S=-5$ & $Q$ & $I$ 
    & $[J_{\rm A},F_{\rm S}]$ & $[J_{\rm S},F_{\rm A}]$ \\
    \hline
    $\Omega^-\Xi^{*-}$ & $-2$ & $1/2$ & $[0,\bm{28}],[2,\bm{28}]$ 
     & $[1,\bm{35}],[3,\bm{35}]$    \\
    $\Omega^-\Xi^{*0}$ & $-1$  &  \\
\end{tabular}
\end{ruledtabular}
\end{table}

\begin{table}[tbp]
\caption{Same with Table~\ref{tbl:S=0 secters} but for the $S=-6$ ($Y=-4$) sector. \label{tbl:S=-6 secters}}
\begin{ruledtabular}
    \begin{tabular}{lcccc}
    Channels for $S=-6$ & $Q$ & $I$ & $[J_{\rm A},F_{\rm S}]$ & $[J_{\rm S},F_{\rm A}]$ \\
    \hline
    $\Omega^-\Omega^-$ & $-2$ & $0$ & $[0,\bm{28}],[2,\bm{28}]$    \\
\end{tabular}
\end{ruledtabular}
\end{table}

\section{Entanglement Power in $nn$ Scattering}\label{app2}

Here we show that the $k$-th-order-weighted entanglement power in Eq.~\eqref{eq:entid} gives the desired result $E_k(\hat{S})=1/2$ for the $nn$ scattering, independent of $k$. For the $nn$ scattering, the $S$-matrix is given by a single component
\begin{align}
   \hat S=\mathcal{J}_{0}\,e^{2i\delta_{0}}\,,
   \label{eq:nnSmatrix}
\end{align}
with $\mathcal{J}_{0}=\mathcal{P}_{\rm A}$.
The projection operator can be written as
\begin{align}
   \mathcal{J}_{0}=\ket{J=0}\bra{J=0},
   \quad 
   \ket{J=0} = \frac{1}{\sqrt{2}}
   \left(\ket{\uparrow\downarrow}
   -\ket{\downarrow\uparrow}\right) .
\end{align}
Let us write the initial state in the computational basis~\cite{Liu:2022grf}
\begin{align}
   \ket{\psi_{\rm in}}
   &=\alpha\ket{\uparrow\uparrow}
   +\beta\ket{\uparrow\downarrow}
   +\gamma\ket{\downarrow\uparrow}
   +\delta\ket{\downarrow\downarrow},
\end{align}
where $\alpha$, $\beta$, $\gamma$, $\delta$ are complex coefficients satisfying $|\alpha|^{2}+|\beta|^{2}+|\gamma|^{2}+|\delta|^{2}=1$. Multiplying the projection operator $\mathcal{J}_{0}$ to the initial state $\ket{\psi_{\rm in}}$, we obtain
\begin{align}
   \mathcal{J}_{0}\ket{\psi_{\rm in}}
   &=\frac{\beta-\gamma}{\sqrt{2}}
   \ket{J=0} .
\end{align}
This means that the normalization factor is
\begin{align}
   \bra{\psi_{\rm in}}\mathcal{P}_{\rm A}\ket{\psi_{\rm in}}
   &=\bra{\psi_{\rm in}}\mathcal{J}_{0}\ket{\psi_{\rm in}} 
   =\frac{|\beta-\gamma|^{2}}{2} \,,
\end{align}
and the normalized density matrix is nothing but the projection operator
\begin{align}
   \tilde \rho 
   & =\frac{\hat{S}_{\rm A}\ket{\psi_{\rm in}} \bra{\psi_{\rm in}} \hat{S}_{\rm A}^{\dagger}}
   {\bra{\psi_{\rm in}}\mathcal{P}_{\rm A}\ket{\psi_{\rm in}}} 
   =\frac{
   \frac{\beta-\gamma}{\sqrt{2}}
   \ket{J=0} 
   \bra{J=0} \frac{(\beta-\gamma)^{*}}{\sqrt{2}}}
   {\frac{|\beta-\gamma|^{2}}{2}} 
   =\mathcal{J}_{0} \,.
   \label{eq:tilderhonn}
\end{align}
Because this normalized density matrix is independent of the angles $\omega_A$ and $\omega_B$ of the initial state, the entanglement power is independent of $k$:
\begin{align}
   E_k(\hat S)
   &=1-\frac{\int\mathrm{d}\omega_A\mathrm{d}\omega_B\langle \psi _{\mathrm{in}}|\mathcal P|\psi _{\mathrm{in}}\rangle^k
   \operatorname{Tr}_A\left[\tilde{\rho}_A^2\right]}
   {\int\mathrm{d}\omega_A\mathrm{d}\omega_B\langle \psi _{\mathrm{in}}|\mathcal P|\psi _{\mathrm{in}}\rangle^k} 
   =1-\operatorname{Tr}_A\left[\tilde{\rho}_A^2\right] .
   \label{eq:EPknn}
\end{align}
Moreover, this is the entanglement entropy of the maximally entangled Bell state, and therefore we find 
\begin{align}
   E_k(\hat S)
   &=\frac{1}{2} \quad (nn \text{ scattering}) \,.
\end{align}
This result is to be expected because the initial state is constrained by FD statistics to be  a maximally entangled Bell state and $\hat S$ is proportional to ${\cal J}_0$, which is an Identity gate over the antisymmetric Hilbert space. So the $S$-matrix preserves the entanglement of the initial state to be 1/2. When averaging over  the antisymmetric Hilbert space, the entanglement power remains at 1/2. Without the proper volume normalization in Eq.~\eqref{eq:EPknn}, the entanglement power would not have come out correctly.

Note that the simple result in Eq.~\eqref{eq:tilderhonn} is obtained because of the single-component $S$-matrix as in Eq.~\eqref{eq:nnSmatrix}. For multi-component $S$-matrix such as Eq.~\eqref{eq:SOmegaOmega} for $\Omega\Omega$ scattering, the normalized density matrix depends on the phase shifts and the angles $\omega_A$ and $\omega_B$, and the cancellation in Eq.~\eqref{eq:EPknn} does not hold. 

\bibliography{refs}

\end{document}